\documentclass[a4paper,onecolumn,oneside,11pt]{article}
\usepackage[utf8]{inputenc}
\usepackage{geometry}
\usepackage{amsmath, amsfonts,amssymb,theorem,
euscript,array,enumerate,amsfonts,mathrsfs,color}
\usepackage{graphics}
\usepackage{epstopdf}
\usepackage[lofdepth,lotdepth]{subfig}

\usepackage{natbib}

\usepackage[dvips]{epsfig}
\input xy
\xyoption{all}

\definecolor{Ying}{rgb}{0.8,0,0.5}

\newcommand{\bde}{\begin{displaymath}}
\newcommand{\ede}{\end{displaymath}}
\newcommand{\el}{\end{lem}}
\newcommand{\be}{\begin{equation}}
\newcommand{\ee}{\end{equation}}
\newcommand{\beq}{\begin{eqnarray*}}
\newcommand{\eeq}{\end{eqnarray*}}
\newcommand{\beqa}{\begin{eqnarray}}
\newcommand{\eeqa}{\end{eqnarray}}
\newcommand{\bel }{\left\{\begin{array}{ll}}
\newcommand{\eel}{\cr \end{array} \right.}
\newcommand{\bex}{\begin{ex} \rm }
\newcommand{\eex}{\end{ex}}

\newcommand{\bp}{\begin{pro}}
\newcommand{\ep}{\end{pro}}

\def\hat{\widehat}

\def\ee{\epsilon}

\def\pt{\partial}

\def\RR{\mathbf{R}}

\def\QQ{\mathbf{Q}}

\def\beqlb{\begin{eqnarray}}\def\eeqlb{\end{eqnarray}}
\def\beqnn{\begin{eqnarray*}}\def\eeqnn{\end{eqnarray*}}

\theoremstyle{plain}
\theorembodyfont{\upshape\it}

\theorembodyfont{\upshape}

\def\edoc{\end{document} }

\providecommand{\keywords}[1]
{
  \small	
  \textbf{\textit{Keywords---}} #1
}

\begin{document}
\thispagestyle{empty}
\author{Bin Xie \footnote{University of Georgia, Department of Mathematics, Athens, GA 30602, 405-438-9933, USA. and Beijing Wuzi University, Tongzhou, Beijing 101149, P.R.China. \texttt{bin.xie@uga.edu}} \and Weiping Li \footnote{Southwest Jiaotong University , Chengdu, Sichuan Province 611756, P.R.China and Oklahoma State University, Stillwater, OK 74078, 405-744-5852, USA. \texttt{w.li@okstate.edu}} \and Nan Liang \footnote{Louisiana State University, Stephenson Department of Entrepreneurship \& Information Systems, Baton Rouge, LA 70803, 225-270-9589, USA. \texttt{nliang@lsu.edu}}}
\date{}
\title{Pricing S\&P 500 Index Options with L\'evy Jumps}

\maketitle


\abstract{}
	We analyze various jumps for Heston model, 
	non-IID model and three L\'evy jump models for S\&P
	500 index options. The L\'evy jump for the S\&P 500 index options is
	inevitable from empirical studies. We estimate parameters from in-sample pricing through SSE for the BS, SV, SVJ, non-IID and L\'evy (GH, NIG, CGMY) models by the method of \cite{bakshi1997empirical}, and utilize them for out-of-sample pricing and compare these models. The  sensitivities of the call option pricing for the L\'evy models with respect to parameters are presented.
	Empirically, we show that the NIG model, SV and SVJ models with estimated volatilities  outperform other models for both in-sample and out-of-sample periods.  Using the in-sample optimized parameters, we find that the NIG model has the least SSE and outperforms the rest models on one-day prediction.


\newpage

\section{Introduction}

Option pricing puzzle has been discussed through plenty of different improvements base on models in \cite{black1973pricing} since 1970's. The most well-known asset returns pricing models include the jump-diffusion model of \cite{merton1976option} and \cite{bates1991crash}; the stochastic volatility model of \cite{heston1993closed}, \cite{hull1987pricing} and \cite{scott1987option}; or both as the stochastic volatility jump-diffusion model of \cite{bates1996jumps}, \cite{scott1997pricing} and \cite{duffie2000transform}. Stochastic interest rate is added in \cite{merton1973theory} and the stochastic volatility and stochastic interest rate are both added in models of \cite{amin1993option}, \cite{bakshi1997alternative}, \cite{scott1997pricing} and \cite{he2018closed}.

\cite{bakshi1997empirical} exhibits empirical results of all those models above for the S\&P 500 call option data by the square of sum of pricing errors (SSE) optimization. They find
the misspecification of all models in different levels.
 The Black-Scholes (BS) model has the most pricing error and the stochastic volatility jump-diffusion (SVJ) model has the least bias. The SVJ model  is the least and the stochastic volatility (SV) model is the second least misspecified models by comparing the SSE values. The stochastic interest (SI) model doesn't improve the BS model significantly and the stochastic volatility, stochastic interest rate and random jump (SVSI-J) model does not enhance SVJ model too much as well. The stochastic volatility contributes the most in the calibration for the period used in \cite{bakshi1997empirical}. However, the volatility is very difficult to verify in the market in fact, and contributions from the compensated Poisson jumps have inherent bias for the random jump process.  \cite{bakshi1997empirical} builds up Heston call option formulas for all previous mentioned models, but the jump assumption is independent identical distributed (i.i.d) Poisson jump which has log-normal distributed jump sizes.
 There are other jumps like non-iid in \cite{camara2008jump} and L\'evy jumps in \cite{Cont2003financial} which might have worth to explore.

 Our paper inspects 27,363 entries as in-sample data and 34,468 entries as out-of-sample data of S\&P 500 option prices. We work on the consistence of SV and SVJ models of \cite{bakshi1997empirical} and find these two models improve the performances of SSE value around 95 present less for both in-sample and out-of-sample comparing BS model. The SVJ models does not reduce bias a lot though the jump part is taken into account, where the jump part in \cite{bakshi1997empirical} is the i.i.d typed compound Poisson jump process. Besides these two models, the non-iid cases of \cite{camara2008jump} which relax the restricts of i.i.d jump distribution are considered but the SSE results improve slightly since the core dynamic does not change too much. 

\cite{constantinides2008mispricing} studies the S\&P 500 index returns, where the underlying asset price is assumed to be the log-normal distribution in BS model without imposing parametric model. Their empirical results according to mispricing suggest that the option market is priced by a different probability distribution beyond what they estimated or modeled. They analyze the one month S\&P 500 index options market from 1986 to 2006, which covers the stock market crash in October, 1987  with seven different samples of index return. Nevertheless the BS model fits the precrash option prices rationally well, the constant volatility setup is not reasonable. The precrash options cannot be priced correctly when the index return distribution is estimated from time series data even by using various statistical adjustments. \cite{broadie2009understanding} disclose the evidence of index option mispricing via option pricing models and stick out the statistical difficulties for option returns analysis. Empirical findings of S\&P 500 put option returns suggest the puzzle that the large average out-of-money put returns is statistically consistent with the BS model, but the average returns, Sharpe ratios and CAPM alphas for deep OTM put returns are insignificant with the BS model statistically. Other evidences show that put returns are inconsistent with the Heston SV model as well.

To explore the jump role in the option pricing dynamics, we try to use L\'evy jump processes to test the mis-specification. L\'evy process is a kind of stochastic process which has independent, stationary increments and jumps.  L\'evy process is an analog of random walk which is the basic simulation of dynamics of market price over time.  L\'evy process can be fully represented by its corresponding characteristic function due to L\'evy-Khinchin theorem. With the characteristic function of the L\'evy process, the expectation integral of European call option pricing
becomes easier to calculate than the according computation of density function of L\'evy process. In recent research, \cite{lee2010detecting} present the signs of small and big jumps of L\'evy processes with reasonable high belief by non-parametric tests. Testing of Dow Jones, S\&P 500 index and individual stocks shows that indices and stocks indeed have different L\'evy jump dynamics. The L\'evy jump evidence is consistent with \cite{li2006bayesian} which study the US market index by using a Bayesian Markov chain Monte Carlo method. \cite{ornthanalai2014levy} studies the contribution of infinitely active L\'evy jumps in the equity risk premium, suggests that kind of L\'evy jumps have dominant role beyond Brownian motions. \cite{zaevski2014option} obtain a general formula for the European option via infinitely divisible L\'evy jumps produced by a tempered stable process, along with statistical test and forecasting. \cite{li2015hedging} disclose the explicit option pricing formula for underlying assets driven by the exponential L\'evy process with gamma jumps. \cite{cont2005integro} introduce the European option pricing as a solution of PIDE (Partial Integro-Differential Equation) for the underlying asset price as exponential L\'evy process and the cases of one or two barriers. The PIDE from L\'evy process model  is very tough to solve explicitly, numerical methods in general include PDE2D scheme of \cite{florescu2014numerical} and fast Fourier transform (FFT) approximation. \cite{duffie2000transform} discuss the analytic utility of Fourier transform in affine models and the main method is FFT estimation of \cite{carr2004time}, \cite{lord2008fast} and \cite{wong2011fft}. \cite{kwok2012efficient} and \cite{hirsa2016computational} elaborate the FFT algorithm which is applied in the calculation of option price with L\'evy process. As an efficiently numerical method to solve  PIDE relevant to L\'evy process, the FFT method may evaluate the option price by interpolation inversely.  Therefore we can check the sensitivities (Greeks) easily from the computation as a by-product.

To estimate the parameters of specific L\'evy process, there are several statistical methods as maximum likelihood estimation (MLE), generalized method of moments (GMM) etc. Nevertheless there is a lack of the closed form of likelihood function of many L\'evy processes. Even for known cases like generalized hyperbolic model, \cite{barndorff1981hyperbolic} and \cite{blaesild1992hyp} point out some flaws of MLE method. The GMM method of \cite{hansen1982large} builds up an estimator for fitting the empirical and theoretical moments to evaluate the parameters. \cite{loretan1994testing} and \cite{lux2000moment} point out that there are some issues of consistency and asymptotic normality of GMM estimator for L\'evy process.
\cite{geman2001time} indicate that the financial assets price processes should include a jump component but it is unnecessary to have a diffusion component. The infinite arriving jump processes dynamics may represent the diffusion component contribution. As \cite{Cont2003financial} also mentioned, L\'evy processes may cause high variability of realized volatility since heavy tailed increments, which means stochastic volatility effects are realized for free. 

In this paper, we use the sum of squared weighted error method of \cite{bakshi1997empirical} and \cite{christoffersen2009shape} uniformly and consistently to estimate parameters for SV, SVJ, non-iid and three special L\'evy processes. Those three L\'evy processes are general hyperbolic (GH) distribution model of \cite{barndorff1977exponentially}, normal inverse Gaussian distribution (NIG) model of \cite{barndorff1997processes} and Carr-Geman-Madan-Yor distribution (CGMY) model of \cite{carr2002fine}. Previous studies on L\'evy models use MLE method to estimate parameters of VG, NIG and CGMY models (see \cite{ornthanalai2014levy}), the different approach to estimate parameters makes it is hard to compare both BS and SV, SVJ models with L\'evy models together. Under the same method with parameters estimated from in-sample S\&P 500 index options, we evaluate and predict the SSE for both in-sample and out-of-sample periods for all models. By comparing BS, SV, SVJ models with different time period in \cite{bakshi1997empirical}, we find the consistent results that the BS model is the least performed, and the SV and SVJ models with estimated volatility perform better than SV and SVJ models with implied volatility model. For both in-sample and out-of-sample, NIG model outperforms the GH and CGMY models consistently. The NIG model, SV and SVJ models with estimated volatility are the best among all the models for in-sample data. The SV and SVJ model with estimated volatility outperform the rest models for the out-of-sample data, and NIG follows next. For one-day prediction, using the in-sample estimated parameters, we find that NIG has the least SSE, and outperforms the rest models on March 1st, 2013 prediction.

The rest of paper is outlined as follows. Section 2 gives data description of S\&P 500 index options. Section 3 introduces the methodology includes the Heston model, non-iid jumps model and L\'evy jump models. Empirical analysis  on S\&P 500 index option is given in Section 4, including in-sample fitting, out-of-sample and one-day prediction.
 Section 5 concludes.

\section{Data Description}

      Due to the most actively traded European style option contracts and data availability, as well as the continuation of the empirical analysis on option pricing models of \cite{bakshi1997empirical}, we use S\&P 500 call option prices for our option pricing analysis.
      
The Standard \& Poor (S\&P) 500 index is a market-capitalization-weighted index of the 500 largest U.S. publicly traded companies. The index is widely regarded as the best gauge of large-cap U.S. equities. It becomes the most popular subject in financial quantitative study since it is representative and data available. 

	The in-sample data are collected from the underlying S\&P 500 index and its call option prices from September 4th, 2012 to February 28th, 2013, and the out-of-sample data are such data from March 1st, 2013 to August 30th, 2013.. The data are obtained from Wharton Research Data Services website (WRDS). The interest rate $r$ is using Treasury Bill Rates from treasury.gov on the corresponding period. We follow steps in \cite{bakshi1997empirical} to filter the data by several criteria:
	
	First, the arbitrage restriction is applied:
\begin{equation}\label{noarb}
C(t,\tau)\geq max(0,S(t)-K,S(t)e^{-q\tau}-Ke^{-r\tau}),
\end{equation}
where $S(t)$ is the underlying S\&P 500 index price, $K$ is the strike price, $\tau$ is time to maturity for the contract and $q$ is the dividend yield.

	Second, the options with less than 6 days to expire are eliminated since they could cause liquidity-related biases.
	
Third, quotes prices of less than \$$3/8$ are excluded because these impacted by volatile bid and ask prices.
	
	After the filtration, 27,363 S\&P 500 index option prices are extracted from the in-sample 176,105 entries. Meanwhile, 34,468 observations are left from the out-of-sample 203,709 items.
	
	Base on the moneyness ratio of $S$ and $K$, we separate these observation as 5 groups: Deep out-of-the-money(DOTM) when $S/K$ is less than 0.94 and out-of-the-money(OTM) as ratio is between 0.94 and 0.97; At-the-money(ATM) when $S/K$ is between 0.97 and 1.03; In-the-money(ITM) if $S/K$ is between 1.03 and 1.06; and deep in-the-money(DITM) if the ratio is beyond 1.06.
	
	Since the database counts the time as working days, which means one year only has 252 days. The options are classified as 3 categories depend on the time to maturity: short-term for less than 40 days, medium-term for 40-120 days, long-term for greater than 120 days\footnote{Bakshi, Cao and Chen (1997) use the sample period from June 1, 1988 to May 31, 1991 with same regions for the moneyness ratio. By the term expiration, they refer to an option contract is (i)short-term ($<60$ days), ours short term ($<40$ days); (ii)medium-term (60--180 days), ours medium-term (40--120 days); (iii) long-term ($>180$ days), ours long-term ($\ge 120$ days)}.
	
	Table \ref{stat1} describes the statistics for these 27,363 in-sample observations. 

\begin{table}[!htbp]
\small
\flushleft
\caption{Statistics for In-sample S\&P 500 Index Call Options
(Sept. 4th 2012-Feb. 28th 2013)}
The sample has 27363 S\&P 500 index call option prices from September 4th 2012 to February 28th 2013 as in-sample. S is the spot S\&P 500 index price and K is the option strike price. DOTM, OTM, ATM, ITM and DITM are deep-out-of-the-money, out-of-the-money, at-the-money, and in-the-money, deep-in-the-money respectively. The statistics in the table respectively are average bid-ask mid-point price, the average effective bid-ask spread (half value of the difference of bid and ask prices),  the average implied volatility and the subtotal number of observations.\\

\label{stat1}

\resizebox{\textwidth}{!}{%
\begin{tabular}{llllll}
\hline
 &  & \multicolumn{3}{l}{Days-to-Expiration} &  \\ \cline{3-5}
Moneyness (S/K) & Statistics & \textless{}40 & 40-120 & $\geq$120 & Total \\ \hline
DOTM (\textless{}0.94) & B-A mid-point price & 0.7436 & 2.3461 & 5.5336 &  \\
 & Eff. B-A Spread & 0.2202 & 0.3512 & 0.5721 &  \\
 & Imp. Volatility & 0.1312 & 0.1231 & 0.1389 &  \\
 & Subtotal & 1292 & 4915 & 6973 & 13180 \\
OTM (0.94-0.97) & B-A mid-point price & 2.0842 & 10.4154 & 28.3255 &  \\
 & Eff. B-A Spread & 0.2385 & 0.5572 & 0.8866 &  \\
 & Imp. Volatility & 0.1232 & 0.1256 & 0.1514 &  \\
 & Subtotal & 3771 & 2492 & 108 & 6371 \\
ATM (0.97-1.00) & B-A mid-point price & 9.0200 & 20.9059 & 44.5500 &  \\
 & Eff. B-A Spread & 0.3502 & 0.6727 & 0.9500 &  \\
 & Imp. Volatility & 0.1248 & 0.1364 & 0.1711 &  \\
 & Subtotal & 5186 & 1167 & 2 & 6355 \\
ATM (1.00-1.03) & B-A mid-point price & 24.4439 & 41.9865 & 0 &  \\
 & Eff. B-A Spread & 0.5747 & 0.8081 & 0 &  \\
 & Imp. Volatility & 0.1610 & 0.1696 & 0 &  \\
 & Subtotal & 1376 & 37 & 0 & 1413 \\
ITM (1.03-1.06) & B-A mid-point price & 63.7440 & 0 & 0 &  \\
 & Eff. B-A Spread & 1.5631 & 0 & 0 &  \\
 & Imp. Volatility & 0.2862 & 0 & 0 &  \\
 & Subtotal & 42 & 0 & 0 & 42 \\
DITM ($\geq$1.06) & B-A mid-point price & 88.7500 & 0 & 0 &  \\
 & Eff. B-A Spread & 1.8000 & 0 & 0 &  \\
 & Imp. Volatility & 0.4146 & 0 & 0 &  \\
 & Subtotal & 2 & 0 & 0 & 2 \\
Total &  & 11669 & 8611 & 7083 & 27363 \\ \hline
\end{tabular}%
}
\end{table}

Table 1 illustrates the in-sample properties of the S\&P 500 index European-style option
prices. Summary statistics are reported for the average bid-ask midpoint price (B-A mid-point price), the average effective bid-ask spread (Eff. B-A Spread), the average of implied volatility (Imp. Volatility)  and the total number of observations for each category.
For each moneyness category, the long term has the maximum effective bid-ask spread and the maximum of implied volatility.
There are plenty of zeros for the long terms in ATM, ITM ans DITM categories, as well as for the medium term in ITM and DITM categories.

Table \ref{stat2} describes the statistics for these 34,468 observations as out-of-sample.

\begin{table}[!htbp]
\small
\caption{Statistics for Out-of-sample S\&P 500 Index Call Options
(Mar. 1st 2013-Aug. 30th 2013)}
\flushleft
The sample has 34468 S\&P 500 index call option prices from March 1st 2013 to August 30th 2013 as out-of-sample. S is the spot S\&P 500 index price and K is the option strike price. DOTM, OTM, ATM, ITM and DITM are deep-out-of-the-money, out-of-the-money, at-the-money, in-the-money, deep-in-the-money respectively. The statistics in the table respectively are average bid-ask mid-point price, the average effective bid-ask spread(half value of the difference of bid and ask prices), the average implied volatility and the subtotal number.\\

\label{stat2}
\resizebox{\textwidth}{!}{%
\begin{tabular}{llllll}
\hline
 &  & \multicolumn{3}{l}{Days-to-Expiration} &  \\ \cline{3-5}
Moneyness (S/K) & Statistics & \textless{}40 & 40-120 & $\geq$120 & Total \\ \hline
DOTM (\textless{}0.94) & B-A mid-point price & 0.6403 & 2.4764 & 10.9504 &  \\
 & Eff. B-A Spread & 0.276 & 0.3904 & 0.9828 &  \\
 & Imp. Volatility & 0.1176 & 0.1135 & 0.1319 &  \\
 & Subtotal & 906 & 4784 & 8210 & 13900 \\
OTM (0.94-0.97) & B-A mid-point price & 1.7718 & 11.4875 & 32.3869 &  \\
 & Eff. B-A Spread & 0.2277 & 0.5695 & 0.9340 &  \\
 & Imp. Volatility & 0.1107 & 0.1174 & 0.1395 &  \\
 & Subtotal & 4172 & 3321 & 798 & 8291 \\
ATM (0.97-1.00) & B-A mid-point price & 9.3417 & 16.3245 & 47.8174 &  \\
 & Eff. B-A Spread & 0.3095 & 0.7656 & 1.0511 &  \\
 & Imp. Volatility & 0.1178 & 0.1291 & 0.1460 &  \\
 & Subtotal & 6323 & 2642 & 92 & 9057 \\
ATM (1.00-1.03) & B-A mid-point price & 28.8327 & 43.3683 & 63.6250 & \\
 & Eff. B-A Spread & 0.6036 & 0.8576 & 0.9250 &  \\
 & Imp. Volatility & 0.1475 & 0.1452 & 0.1513 &  \\
 & Subtotal & 2748 & 401 & 2 & 3151 \\
ITM (1.03-1.06) & B-A mid-point price & 63.9500 & 0 & 0 &  \\
 & Eff. B-A Spread & 1.2355 & 0 & 0 &  \\
 & Imp. Volatility & 0.2432 & 0 & 0 &  \\
 & Subtotal & 69 & 0 & 0 & 69 \\
DITM ($\geq$1.06) & B-A mid-point price & 0 & 0 & 0 &  \\
 & Eff. B-A Spread & 0 & 0 & 0 &  \\
 & Imp. Volatility & 0 & 0 & 0 &  \\
 & Subtotal & 0 & 0 & 0 & 0 \\
Total &  & 14218 & 11148 & 9102 & 34468 \\ \hline
\end{tabular}%
}
\end{table}

Table \ref{stat2} illustrates the out-of-sample properties of the S\&P 500 index European-style option prices. 
Summary statistics are reported (similar to Table 1), where the short term total observations are 11,669 and 14,218 respectively for the in-sample and the out-of-sample periods, the medium term total observations are 8,611 and 11,148, the long term total observations are 7,083 and 9,102 respectively for two different periods. For the deep out-of-the-money, the middle term has the minimum average implied volatility for both in-sample and out-of-sample. Constantly, minimum average implied volatility happens at the short term for the other categories.

\section{Methodology}
The purpose of this section is to set our testing models for the jump-diffusion processes in our empirical studies of S\&P 500 index options.
Merton (1976) introduced the first jump processes in option pricing model, where the jump is a Poisson process with constant intensity independent of the
continuous Brownian motion and i.i.d jumping size random variables independent from the Brownian motion and the Poisson jump process. We set the Black-Scholes model as a benchmark for the option pricing model, and first including the Heston model with stochastic volatility under the risk-neutral measure. Then we include the Non-iid jump model to vary the jumping size variables with different means and different variances. Following \cite{camara2008jump}, it is convenient to follow a standard numerical analysis to estimate the option prices under a risk-neutral probability measure. This would naturally lead to consider more general jump processes as L\'evy jump models, and to empirically test three typical L\'evy jump processes (General hyperbolic distribution, Normal Inverse Gaussian distribution, and Carr-Geman-Madan-Yor class of distribution). 

\subsection{Model Dynamics}
In the last four decades, many modifications were made to relax the restrictions of the Black-Scholes (BS) model in \cite{black1973pricing}. Heston (1993) model is a solid way by setting the volatility as a stochastic process as well.
Heston (1993) assumes that the underlying stock price $S_t$ follows the geometric Brownian motion just as the process in the BS model, and
the variance $v_t$ follows a stochastic  CIR process in \cite{cox1985theory}. Then the Poisson jump model has the bi-variate system of Stochastic Differential Equations (SDEs) listing as (see \cite{bakshi1997empirical} and \cite{rouah2013heston}):

\begin{align}
\frac{dS_t}{S_t}  = &(\mu - \lambda \mu_J)  dt +\sqrt{v_t} dW_1(t)+ dq(t)\\
dv_t  = &\kappa (\theta -v_t) dt + \sigma \sqrt{v_t}dW_2(t),
\end{align}
where $\mu$ is the drifting term for the geometric Brownian motion, $\kappa>0$ is the mean reversion speed for the variance, $\theta>0$ is the long-run mean, $\sigma>0$ is the volatility of the variance, $W_1(t)$ and $W_2(t)$ are two Brownian motions with correlation coefficient $\rho $ $(-1 \leq \rho \leq 1)$, $\lambda$ is the frequency of jump per year and $J(t)$ is the percentage jump size which is log-normally distributed over time with unconditional mean $\mu_J$ such that $S(t) = X(t)J(t)$ with the continuous stochastic process part $\frac{dX(t)}{X(t)}=(\mu - \lambda \mu_J)  dt +\sqrt{v_t} dW_1(t)$, $q(t)$ is a Poisson jump counter with intensity $\lambda$. Hence, the parameters for the Heston model as $\lambda =0$ are 
$(\mu, \kappa, \theta, \sigma, \rho)$, and the parameters for the SVJ model 
as $\lambda \neq 0$ consists of those of Heston models.

 If $\sigma=0$ and $\kappa=0$, then the Heston model is reduced to the Black-Scholes model with $\sigma_{BS}=\sqrt{v_t}$. In the afterwards empirical study, we deal with the $v_t$ value in two choices: using the square of implied volatility given in the market database or estimate $v_t$ as a parameter. We implement every model by backing out, on each day, the implied volatility and structural parameters from the market option prices of that day. This is quite a common approach in literature (e.g., \cite{bakshi1997empirical}), to resolve internal consistency on parameters for models we test.

By the Esscher transform, under an risk-neutral measure $\QQ$, the new process SDEs are:
\begin{align}\label{HestonQ}
\frac{dS_t}{S_{t^-}}  = &(r  - \tilde{\lambda}\tilde{\mu_J}) dt +\sqrt{v_t} d\tilde{W}_1(t)+ dq(t),\\
dv_t  = &\kappa (\theta -v_t) dt + \sigma \sqrt{v_t}d\tilde{W}_2(t),
\end{align}
where $d\tilde{W}_1(t)=dW_1(t) + \int_0^t\theta_s ds$ and $\tilde{W}_2(t)= W_2(t)$ are two Brownian motions under the risk-neutral measure $\QQ$ with the correlation $\rho$, and the market price of risk equation
\[\mu - \lambda \mu_J = (r  - \tilde{\lambda}\tilde{\mu_J})+ \sqrt{v_t} \theta_t.\]
If the stock pays a continuous dividend yield $q$, we may replace $r$ in the equation (\ref{HestonQ}) as $r-q$. See \cite{duffie2000transform}, \cite{Cont2003financial} and \cite{shreve2004stochastic} for more details.

The close-form option pricing formula for Heston model SDEs is given by the solution of a difference between two expressions of probability, just as the close-form option pricing formula in the BS model.
For a European call option with a strike price $K$  and time-to-maturity $\tau$, the close-form option  price $C(t,\tau)$ subject to $C(t,\tau)=max\{S(t,\tau)-K,0\}$ is given by
\begin{equation}\label{HestonC}
C(t,\tau)=S(t)\Pi _1 (t,\tau; S, R, V)-KB(t,\tau) \Pi _2 (t,\tau; S, R, V).
\end{equation}
where the two probabilities $\Pi_1$ and $\Pi_2$ are represented by the respective characteristic functions $f_j$'s ($j=1,2$) for the stochastic-volatility jump-diffusion (SVJ) model (see \cite{heston1993closed}, \cite{bates1996jumps}, \cite{scott1997pricing}, \cite{bakshi1997empirical} and \cite{duffie2000transform}):
\begin{equation} \label{HestonPi}
\Pi_j (t,\tau;S_t,R_t,V_t))=\frac{1}{2}+\frac{1}{\pi}\int^\infty_0 
Re[\frac{e^{-i\phi lnK}f_j(t,\tau,S_t,R_t,V_t;\phi)}{i\phi}]d\phi,\ \ \
 j=1,2
\end{equation}

\begin{align*}
f_1= & exp\large \{ -i\phi lnB(t,\tau)-\frac{\theta_v}{\sigma^2_v}[2ln(1
-\frac{[\xi_v-\kappa_v+(1+i\phi)\rho \sigma_v] (1-e^{-\xi_v\tau})}{2\xi_v})]\\
& -\frac{\theta_v}{\sigma^2_v}[\xi_v-\kappa_v+(1+i\phi)\rho \sigma_v]\tau+i\phi lnS(t)]\\
& +\lambda(1+\mu_J)\tau [(1+\mu_J)^{i\phi}e^{(i\phi/2)(1+i\phi)\sigma^2_J}-1]-\lambda i\phi \mu_J\tau\\
& +\frac{i\phi(i\phi+1)(1-e^{-\xi_v \tau)}V_t}{2\xi_v-[\xi_v-\kappa_v 
  +(1+i\phi) \rho \sigma_v](1-e^{-\xi_v\tau})}\large\},\\ 
f_2= & exp\large\{-i\phi lnB(t,\tau)-\frac{\theta_v}{\sigma^2_v}[2ln(1
-\frac{[\xi^*_v-\kappa_v+i\phi\rho \sigma_v] (1-e^{-\xi^*_v\tau})}{2\xi^*_v})]\\
& -\frac{\theta_v}{\sigma^2_v}[\xi^*_v-\kappa_v+i\phi\rho \sigma_v]\tau+i\phi lnS(t)]\\
& +\lambda \tau [(1+\mu_J)^{i\phi}e^{(i\phi/2)(i\phi-1)\sigma^2_J}-1]-\lambda i\phi \mu_J\tau\\
& +\frac{i\phi(i\phi-1)(1-e^{-\xi^*_v \tau)}V_t}{2\xi^*_v-[\xi^*_v-\kappa_v 
  +i\phi \rho \sigma_v](1-e^{-\xi^*_v\tau})}\large\},
\end{align*}
where setting $R_t=R$ as a constant risk-free rate and $B(t,\tau)= e^{-r\tau}$ in (\ref{HestonC}), and 
\begin{align*}
\xi_v &=\sqrt{[\kappa_v-(1+i\phi)\rho\sigma_v]^2-i\phi(i\phi+1)\sigma^2_v},\\
\xi^*_v &=\sqrt{[\kappa_v-i\phi\rho\sigma_v]^2-i\phi(i\phi-1)\sigma^2_v}.
\end{align*}

The stochastic volatility (SV) model can be obtained by setting $\lambda=0$, and the SVJ model is for the general nonzero $\lambda$.
\cite{bakshi1997empirical} sets more general SVSI-J model along with five models on the Black-Scholes (BS) model, the stochastic volatility (SV), the stochastic interest-rate (SI or SVSI) and the stochastic volatility random jump (SVJ) to conduct a more comprehensive empirical study on the relative merits of competing option pricing models. They showed that the SI and SVSI-J models do not significantly improve the performance of the BS and SVJ models, based on the S\&P 500 index call options from June 1988 to May 1991. Moreover their empirical findings indicate that the SVJ model has the least misspecified option pricings and the BS has the most.
But the jumping process in the SVJ model is the compound Poisson jump with i.i.d jump size. Hence we first extend the SVJ model to the Non-iid jump model
developed in \cite{camara2008jump}.

\subsection{Non-iid Jump Model}

There is an important assumption of jumps diffusion model initially set up by \cite{merton1976option} as we discussed before is that the underlying asset has the identically and independently distributed (i.i.d) jumps in the compound Poisson process. \cite{camara2008jump} derives the close-form of option pricing formula for the non-iid jumping sizes in the compound Poisson process.  In order to compare with the SVJ and the SV models under the consistent parameters estimation, we analyze three types of the non-iid jumping sizes which are (i) with time-varying means, (ii) with time-varying variances and (iii) with auto-correlated jumps.


By building the non-iid jumps, \cite{camara2008jump} extend a fundamental formula base on the jump-diffusion pricing model in \cite{merton1976option}. 
We summarize their results in the following, in order to estimate and calculate the option prices numerically.

The jump-diffusion call option pricing $P_c$ without i.i.d assumption on jumps is given by:
\begin{equation} \label{Li15} 
P_c= \frac{e^{-\lambda T}}{K_1}\sum^\infty_{n=0} \frac{(\lambda'_n T)^n}{n!}(S_0 
e^{-qT}N(d_{1,n})-Ke^{-r_n T}N(d_{2,n})),
\end{equation}
where

\begin{align*}
d_{1,n} &=\frac{ln(\frac{S_0}{K})+(r_n-q+\frac{\sigma^2_n}{2})T}{\sigma_n \sqrt{T}}, \hspace{.25in}
d_{2,n}=d_{1,n}-\sigma_n \sqrt{T},
\hspace{.25in} K_1 = e^{-\lambda T}\sum^\infty_{n=0} \frac{(\lambda'_n T)^n}{n!}.
\end{align*}


(i) \textbf{Corollary 1 (Jumps with time-varying means}) 

Let $Y_i \sim N(\alpha_i,\gamma^2)$ and $Cov(Y_i,Y_j) =\gamma_{ij} =0 $ $(i\neq j)$. The specific parameters in formula (\ref{Li15}) are computed as 
\begin{align*}
\lambda'_n &=\lambda e^{\bar{\alpha}_n+\gamma^2/2},\text{    }
\sigma^2_n=\frac{n}{T} \gamma^2 + \sigma^2 , \hspace{.2in}
r_n = r -\frac{ln(K_1)}{T}+\frac{n}{T}(\bar{\alpha}_n+\frac{\gamma^2}{2}),\\
\bar{\alpha}_n &= \sum^n_{i=1} \frac{\alpha_i}{n}, \hspace{.25in}
\bar{\alpha}_0=0. 
\end{align*}

(ii) \textbf{Corollary 2 (Jumps with time-varying variances})

Let $Y_i \sim N(\alpha,\gamma^2_i=\gamma_{ii})$ and $Cov(Y_i,Y_j)=0$ $(i\neq j)$. Then the parameters in formula (\ref{Li15}) are given as:
 
\begin{align*}
\lambda'_n &=\lambda e^{\alpha+\bar{\gamma}^2_{n}/2}, \hspace{.2in}
\sigma^2_n=\frac{n}{T} \bar{\gamma}^2_n + \sigma^2, \hspace{.2in}
r_n = r -\frac{ln(K_1)}{T}+\frac{n}{T}(\alpha_n+\frac{\bar{\gamma}^2_n}{2}), \\
\bar{\gamma}^2_n &=\sum^n_{i=1} \frac{\gamma_i^2}{n},
\hspace{.25 in}
\bar{\gamma}^2_0=0.
\end{align*}

(iii) \textbf{Corollary 3 (Autocorrelated Jumps})

Let $Y_i \sim N(\alpha,\gamma^2)$ and $Cov(Y_i,Y_l)=\gamma^2\rho_{il}$. Then we have the parameters in formula (\ref{Li15}) modified as:

\begin{align*}
\lambda'_n &=\lambda e^{\alpha+\frac{\gamma^2}{2}[1+(n-1)\bar{\rho}_n]},
\hspace{.2in}
\sigma^2_n =\sigma^2 +\frac{n}{T} \gamma^2[1+(n-1)\bar{\rho}_n], \\
r_n &= r -\frac{ln(K_1)}{T}+\frac{n}{T}(\alpha+\frac{\gamma^2}{2}[1+(n-1)\bar{\rho}_n]),\\
\bar{\rho}_n &=\sum^n_{i=1,i \neq l}  \sum^n_{l=1} \frac{\rho_{il}}{n(n-1)}, \hspace{.15 in}
\bar{\rho_0}=0,\hspace{.15 in}
\bar{\rho}_1=0. \\
\end{align*}


The formula for the first case is similar to the classic Merton's jump-diffusion pricing formula with modified parameters from the non-iid jump sizes in the jump-diffusion pricing model. Ordinarily, the call price formula here is a function of the stock price $S_0$, strike price $K$, time of maturity $T$, interest rate $r$, stock volatility $\sigma$, Poisson process intensity $\lambda$. The notation $\gamma^2$ is covariances $Cov(Y_i,Y_i)=\gamma_{ii}$ which are fixed, and $\gamma_{ij} =0$ ($i\neq j$) that means jumps affect the time-varying means only. There are some other parameters as $K_1$ is the total gross stock return factor, $\bar{\alpha}_n$ is the moving average for the price jump sizes means.
The second non-iid case may occur on the variances of the price jumps only. The means are fixed as $\alpha$ which is the average values for all $\alpha_i$'s in the first case. The covariance $Cov(Y_i,Y_i)=\gamma^2_{i}$, the moving average of jumps variances $\bar{\gamma}_n^2$ is the new average value which is different with the fixed $\gamma$ value in (i).
The third case of the non-iid jump model which is only on the autocorrelations of price jumps with nontrivial covariaces of jumps.
Unlike the cases (i) and (ii), there is a new factor $\bar{\rho}_n$ as the moving average of the autocorrelations. The moving average for the price jump sizes means $\bar{\alpha}_n$ in (i) and the moving average of jumps variances $\bar{\gamma}_n^2$ in (ii) are constants. See \cite{camara2008jump} for more details.

\subsection{L\'evy Jump Models}

L\'evy process enables flexible modelling of the distribution of returns at a given time horizon, especially when it comes to model the tails of the distribution. The distribution of asset returns in financial markets seems to carry a heavy tail which has positive excess kurtosis with a tail index, and the
log-price of assets shows stylized empirical properties as the absence of autocorrelation in increments,  heavy tails, finite variance, aggregational normality, jumps in price trajectoies which are natural properties L\'evy processes endow, especially like the variance gamma, NIG, hyperbolic, CGMY L\'evy processes carry the heavy tails.
There are tremendous literature on L\'evy processes  in financial quantitative analysis (see \cite{Cont2003financial}).

In the BS model, the dynamics of asset price is described as an exponential of Brownian motion with drifting term, as the geometric Brownian motion,
$$ S_t= S_0 exp(\mu t +\sigma W_t)= S_0 exp(B_t) ,$$
where $B_t =\mu t +\sigma W_t$.
For the exponential L\'evy process,  we have $S_t= S_0expX_t$, where $X_t$ is a L\'evy process. By the Ito formula and L\'evy-Khinchin decomposition theory, we have the Martingale-drift decomposition of functions.

\cite{Cont2003financial} show that the geometric L\'evy process $Y_t=exp(X_t)$ is a semimartingale with a decomposition $Y_t =M_t+A_t$ of a martingale process and continuous finite variation drifting term,
where $(X_t)$ is a L\'evy process with L\'evy triplet ($\sigma^2,\gamma,\nu$) satisfying
$ \int_{|y|\geq 1}e^y \nu (dy) < \infty,$
and $M_t$ is the martingale part
$$ M_t = 1 + \int^t_0 Y_{s-}\sigma dw_s +\int_{[0,t]\times \RR} Y_{s-}(e^z-1)\tilde{J}_X (ds dz),$$ 
and the continuous finite variation drift part is 
$$A_t = \int ^t_0 Y_{s-}[\gamma +\frac{\sigma^2}{2}+\int_{\RR}(e^z-1-z1_{|z|\leq 1}) \nu (dz)] ds.$$
Therefore, $(Y_t)$ is a martingale if and only if 
$$\gamma +\frac{\sigma^2}{2}+\int_{\RR}(e^z-1-z1_{|z|\leq 1}) \nu (dz)=0.$$


With the L\'evy jump model, the European option price can be evaluated by solving
an Partial Integro-Differential Equation (PIDE) from the martingale property.
Suppose the European option with maturity $T$ and payoff $H(S_T)$  satisfies Lipschitz condition $|H(x)-H(y)|\leq c|x-y|$ for some $c>0$. Hence, the option price under the geometric  L\'evy process asset is given by
\[C(t,s)=e^{-r\tau}E[H(S_T)|\mathcal{F}_t]=e^{-r\tau}E[H(S_t e^{r\tau + X_{\tau}})].\]
Then the risk-neutral dynamic of $S_t$ is obtained by (\cite{Cont2003financial})
\begin{align*}
S_t = & S_0 + \int^t_0 rS_{u-}du +\int^t_0 S_{u-}\sigma dW_u
	  + \int^t_0 \int_{\RR} (e^x-1)S_{u-}\hat{J}_X(du dx),
\end{align*}
where $\hat{J}_X$ is the compensated jump measure of the L\'evy process $X$ and $\hat{S}_t = e^{X_t}$ is a martingale:
$ \frac{d\hat{S}_t}{\hat{S}_{t-}}= \sigma dW_t +\int_{\RR} (e^x-1)\hat{J}_X (dt dx),$
verifies $\sup_{t\in [0,T]}E[\hat{S}^2] <\infty$.
Suppose $\int_{|y|\geq 1}e^{2y} \nu (dy) < \infty.$ If either the volatility $\sigma>0$ strictly positive or 
there exists a $\beta \in [0,2]$ such that $ \lim \limits_{\epsilon \to 0^+} \epsilon ^{-\beta} \int ^{\epsilon}_{-\epsilon}|x|^2 d\nu (x) >0,$
then the option price $C(t,s) :[0,T]\times (0,\infty) \rightarrow \RR$ of a European option with terminal payoff $H(S_T)$ is a solution of the partial integro-differential equation (Backward PIDE for European option with L\'evy process, see \cite{Cont2003financial}):
\begin{equation} \label{PIDE}
\begin{split}
\frac{\pt C}{\pt t}(t,S) &+r\frac{\pt C}{\pt S}(t,S)+\frac{\sigma^2S^2}{2} \frac{\pt^2 C}{\pt^2 S}(t,S)-rC(t,S)\\ 
	&+\int \nu (dy)[C(t,Se^y)-C(t,S)-S(e^y-1)\frac{\pt C}{\pt S}(t,S)]=0
\end{split}
\end{equation}
on $[0,T]\times (0,\infty) $ with the terminal condition
$C(T,S)=H(S).$


By analyzing how to solve the PIDE (\ref{PIDE}), one realized that it is extremely difficult to get the closed-form option pricing formula since
(i) it is a backwards partial differential equation with the boundary condition of maturity time, and (ii) the integral part in the equation is very tough to handle. Fortunately, we can use the Fourier transform to estimate the solution numerically.
Fast Fourier transforms are used in modern applications in engineering, science, and mathematics widely. The Cooley-Turkey FFT algorithm which was introduced in \cite{cooley1965algorithm} can reduce the $N^2$ multiplication of discrete Fourier transform to $NlnN$ by using a divide. \cite{strang1994wavelets} described the FFT in 1994 as "the most important numerical algorithm of our lifetime" and it is included in Top 10 Algorithms of 20th Century by the IEEE journal. For the details how to apply FFT method to estimate the solution, see \cite{kwok2012efficient}, \cite{kienitz2012financial} or  \cite{hirsa2016computational}.

\subsection{Three Typical L\'evy Processes in Finance and Numerical Sensitivity Studies}
Three typical L\'evy processes are specified in order to evaluate the option pricing model explicitly, and the explicit density function and characteristic functions are listed in this subsection for later calibrations. Therefore the explicit formulas of FFT algorithm are determined. The sensitivities for the parameters through numerical solutions are also discussed.

\vspace{.15 in}
\textbf{1. General Hyperbolic (GH) Distribution}

General Hyperbolic distribution is a class of Lebesgue continuous infinitely divisible distribution of 4 parameters. The Lebesgue density is defined in \cite{barndorff1977exponentially}:

\begin{align*}
    \rho_{GH} (x) &= a(\alpha, \beta, \delta, \nu)e^{\beta x}
        (\delta^2 + x^2)^{(\nu - 1/2)/2} K_{\nu-1/2}(\alpha \sqrt{\delta^2+x^2}),\\
       a(\alpha, \beta, \delta, \nu)&= \frac{(\alpha^2-\beta^2)^{\nu /2}}{\sqrt{2 \pi} \alpha ^{\nu - 1/2}\delta^{\nu} K_{\nu}(\delta \sqrt{\alpha^2-\beta^2})},
\end{align*}
where the domain of parameters is specified by $\alpha>0,\beta\in(-\alpha,\alpha),\delta>0$, $\nu\in \RR$,  $K_{\nu}$ and $K_{\nu-1/2}$ are the modified third kind of Bessel functions with the order as subscripts.
The characteristic function of this distribution law is given by
$\phi(u)= (\frac{\alpha^2-\beta^2}{\alpha^2-(\beta+iu)^2})^\frac{\nu}{2}
\frac{K_{\nu}(\delta \sqrt{\alpha^2-(\beta+iu)^2})}{K_{\nu}(\delta \sqrt{\alpha^2-\beta^2)}}.$ When all data are on the same time scale, the sample from a GH model and estimating its parameters are relatively easy. 

To analyze the sensitivities of the parameters in these three kinds of L\'evy processes, we need to pick up some initial parameters (in related literature) to start the SSE procedure as in \cite{bakshi1997empirical}. Here we fulfill the sensitivities investigation base on the optimal parameters of \cite{schoutens2003levy}, which focus on the S\&P 500 index call option prices study by minimizing the root-mean-square error for the price differences of market and models.

\begin{figure}[htbp!]
\centering

\subfloat[Subfigure 1 list of figures text][Call prices - $S_t$]{
\includegraphics[width=0.3\textwidth]{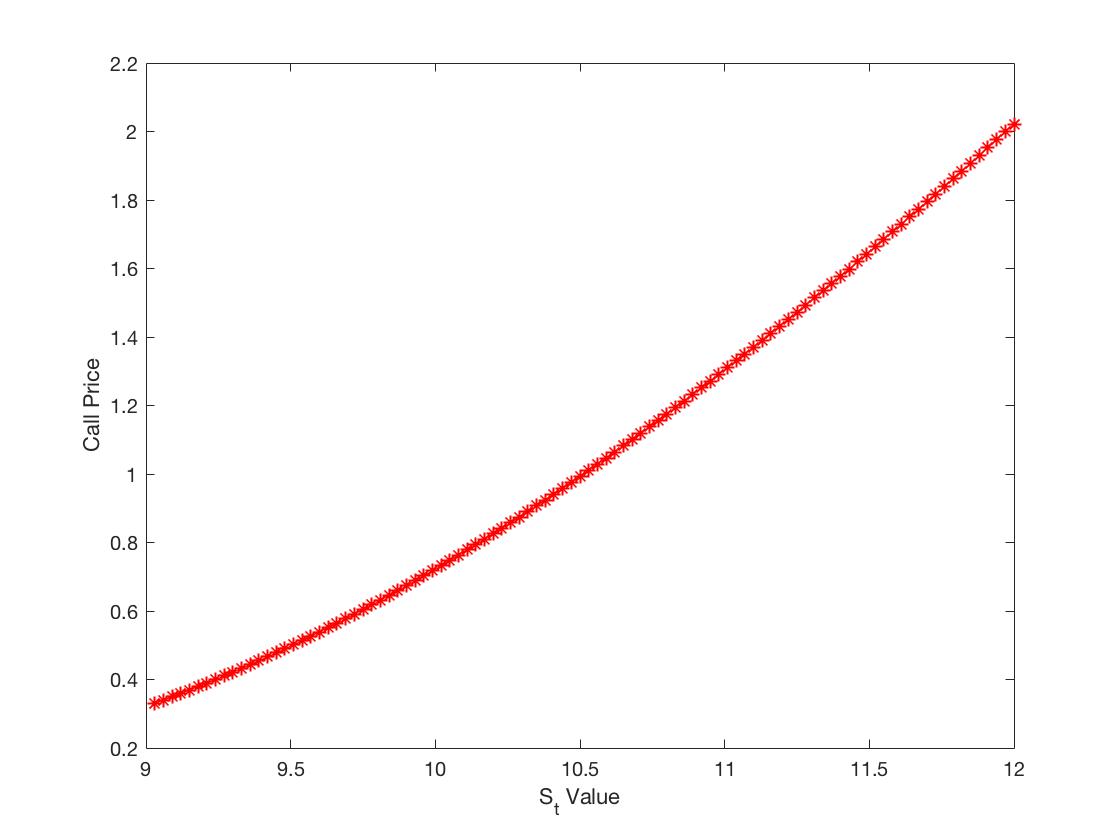}
\label{fig:GHtoS}}
\subfloat[Subfigure 2 list of figures text][Call prices - K]{
\includegraphics[width=0.3\textwidth]{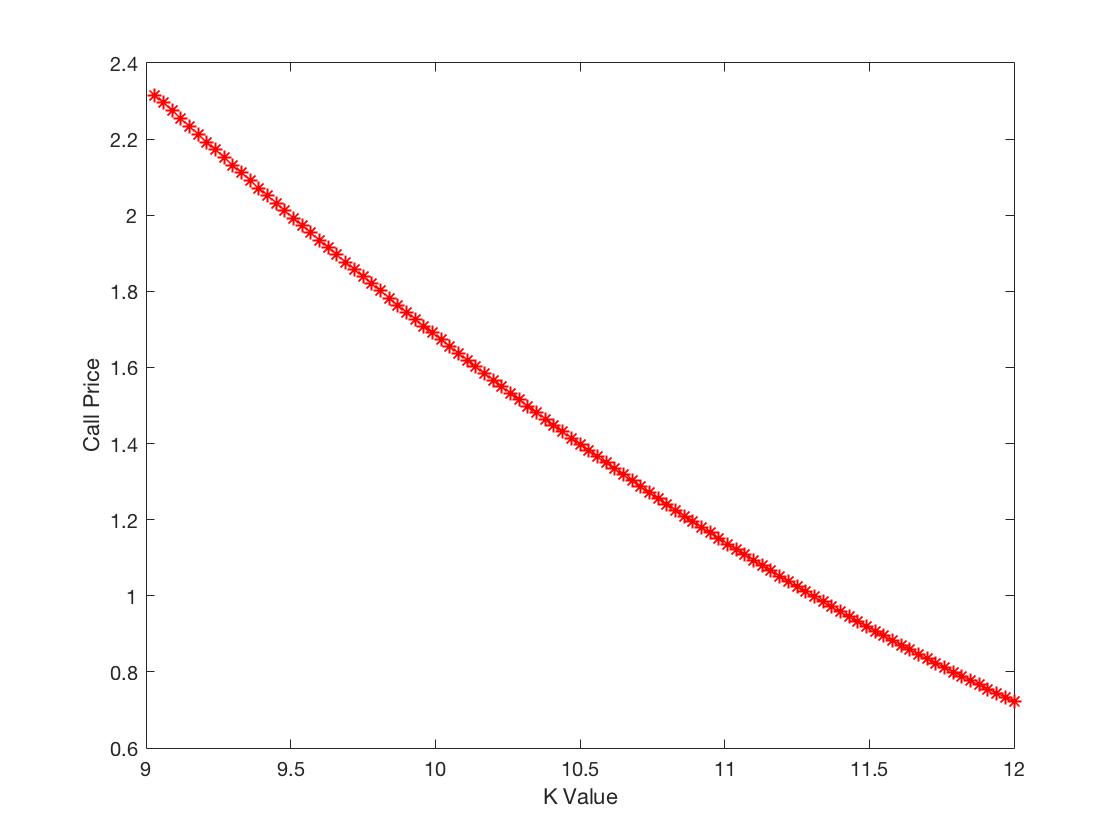}
\label{fig:GHtoK}}
\subfloat[Subfigure 3 list of figures text][Call price - T]{
\includegraphics[width=0.3\textwidth]{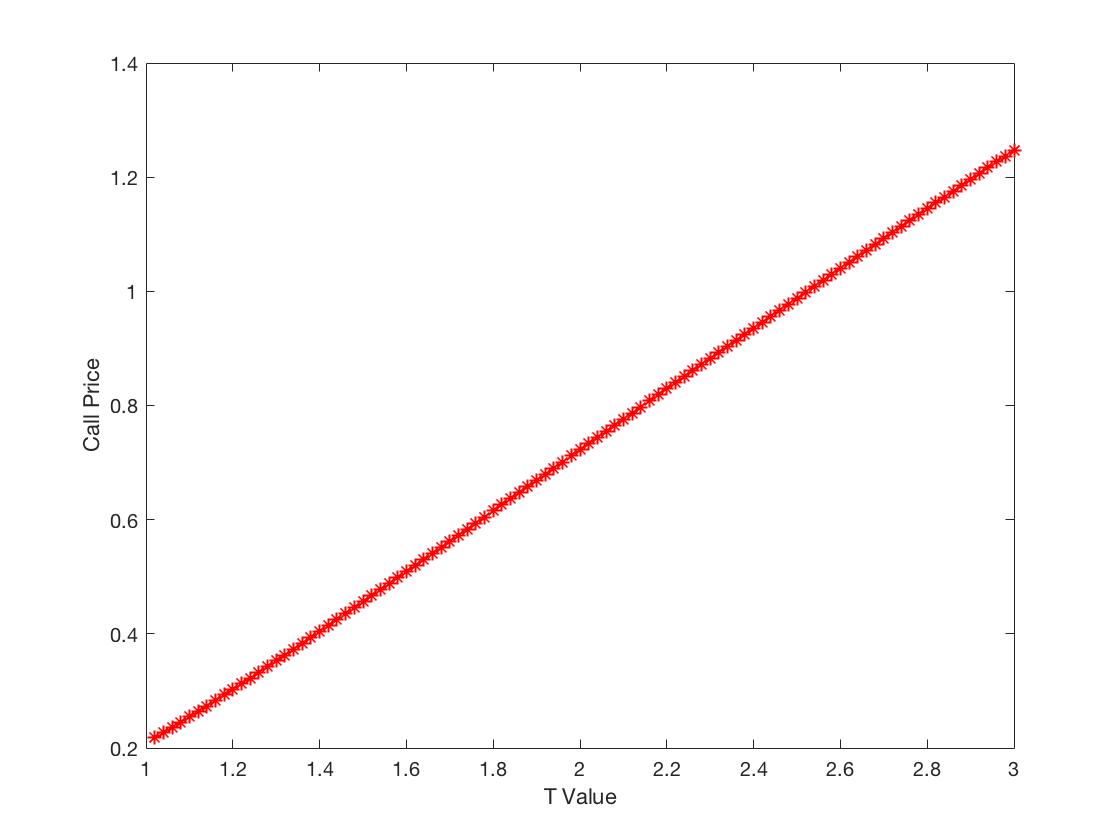}
\label{fig:GHtoT}}

\subfloat[Subfigure 4 list of figures text][Call price - r]{
\includegraphics[width=0.3\textwidth]{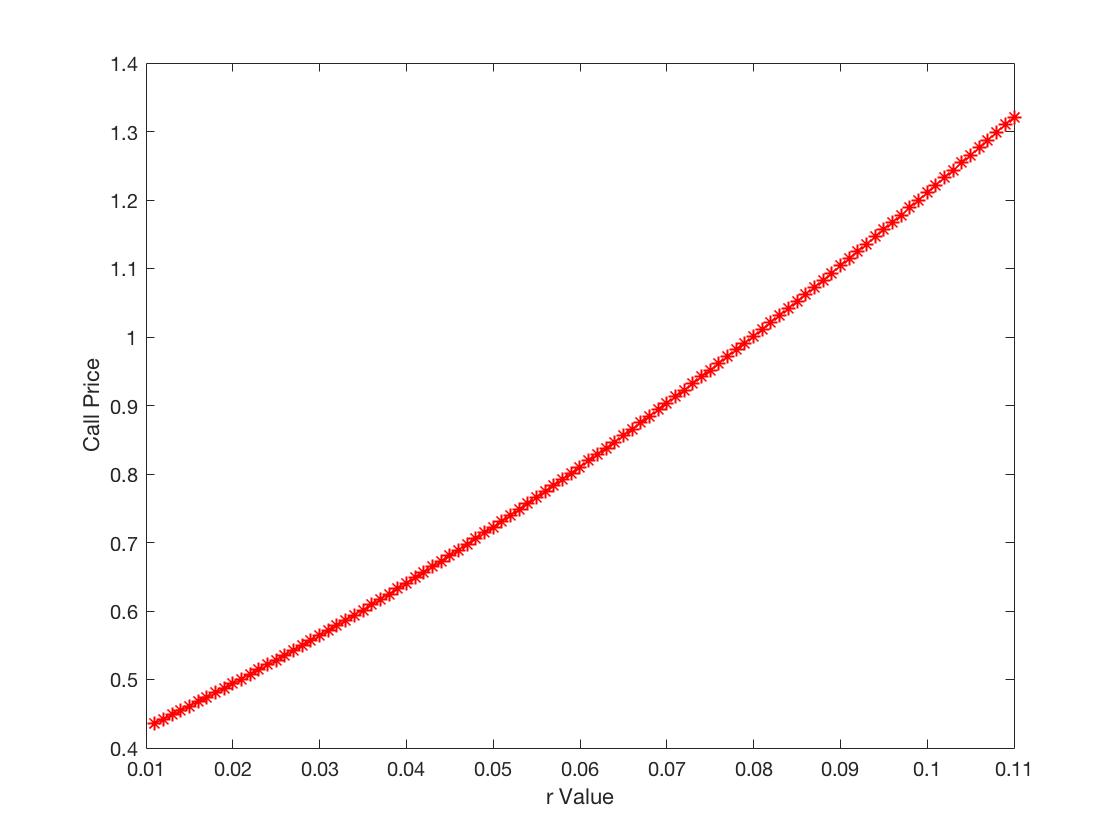}
\label{fig:GHtor}}
\subfloat[Subfigure 5 list of figures text][Call price - q]{
\includegraphics[width=0.3\textwidth]{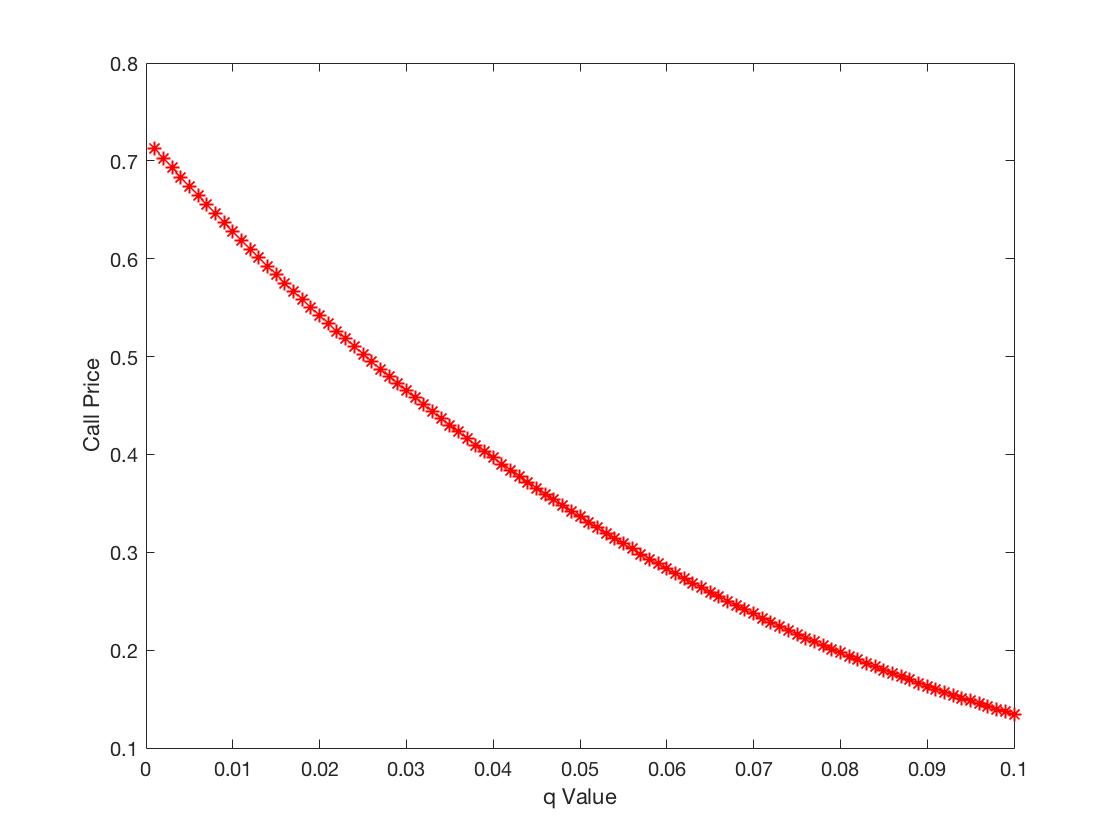}
\label{fig:GHtoq}}
\subfloat[Subfigure 6 list of figures text][Call price - $\alpha$]{
\includegraphics[width=0.3\textwidth]{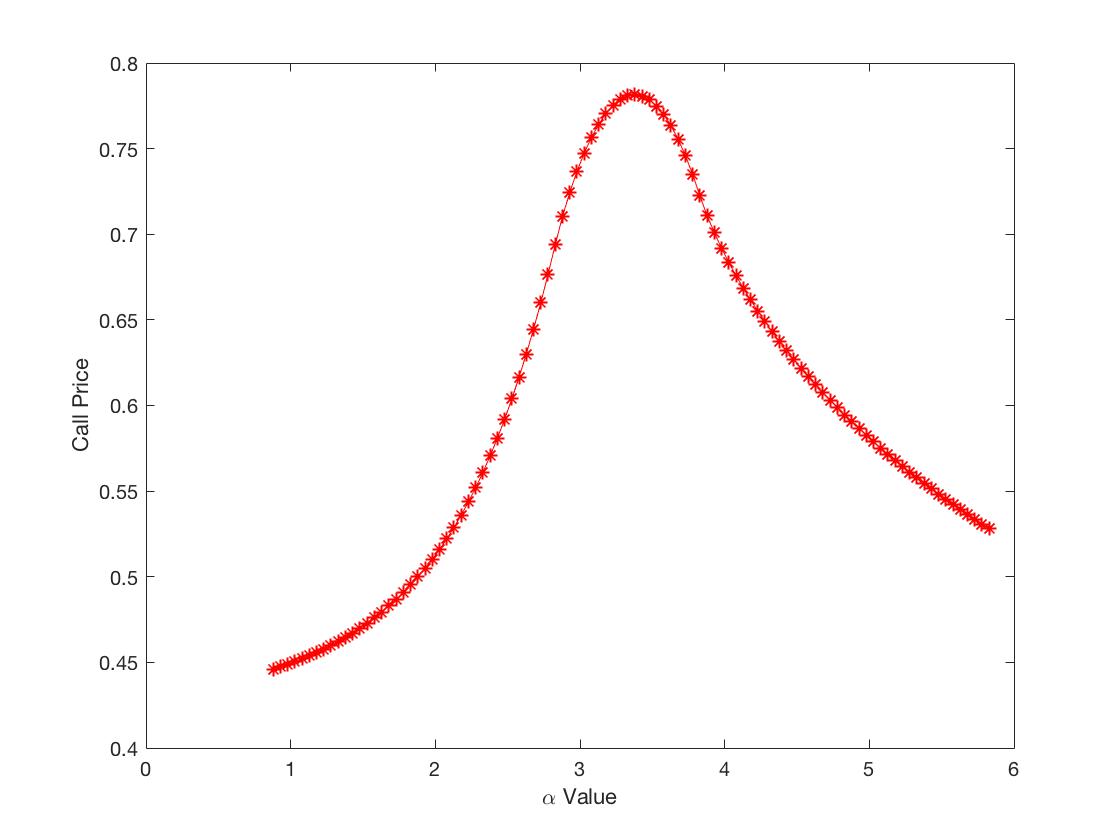}
\label{fig:GHtoalpha}}

\subfloat[Subfigure 7 list of figures text][Call price - $\beta$]{
\includegraphics[width=0.3\textwidth]{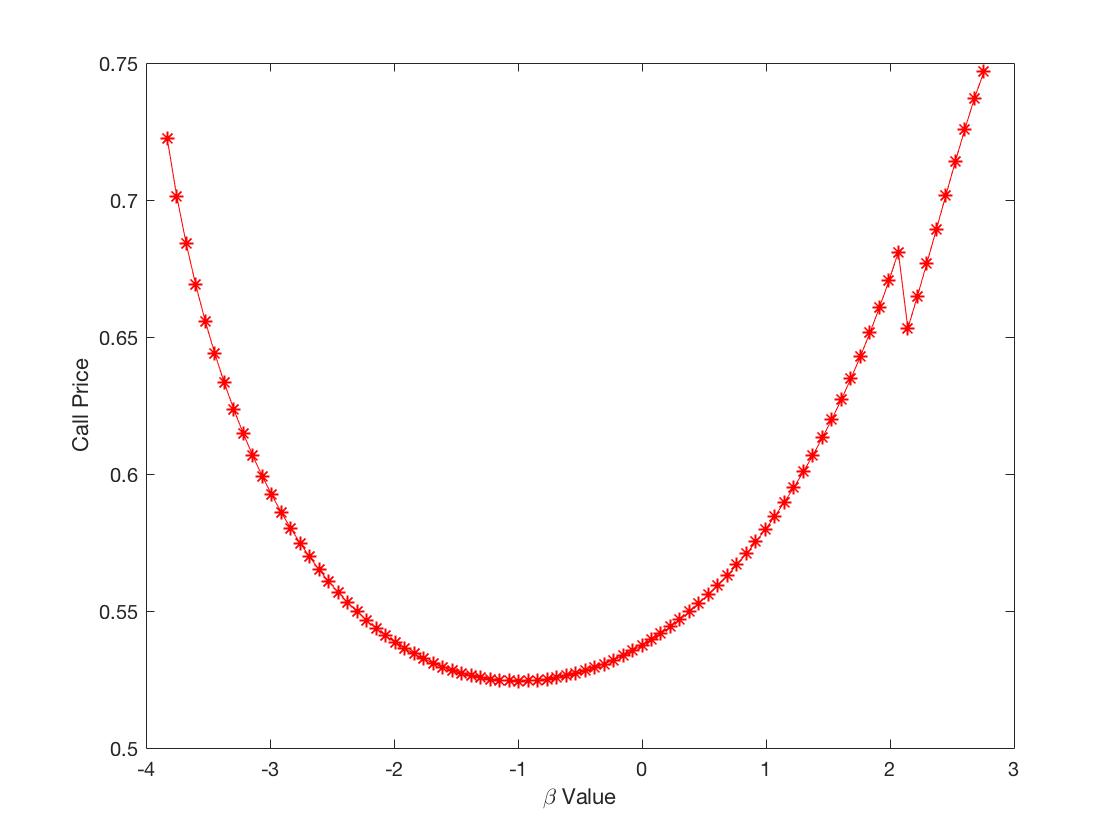}
\label{fig:GHtobeta}}
\subfloat[Subfigure 8 list of figures text][Call price - $\delta$]{
\includegraphics[width=0.3\textwidth]{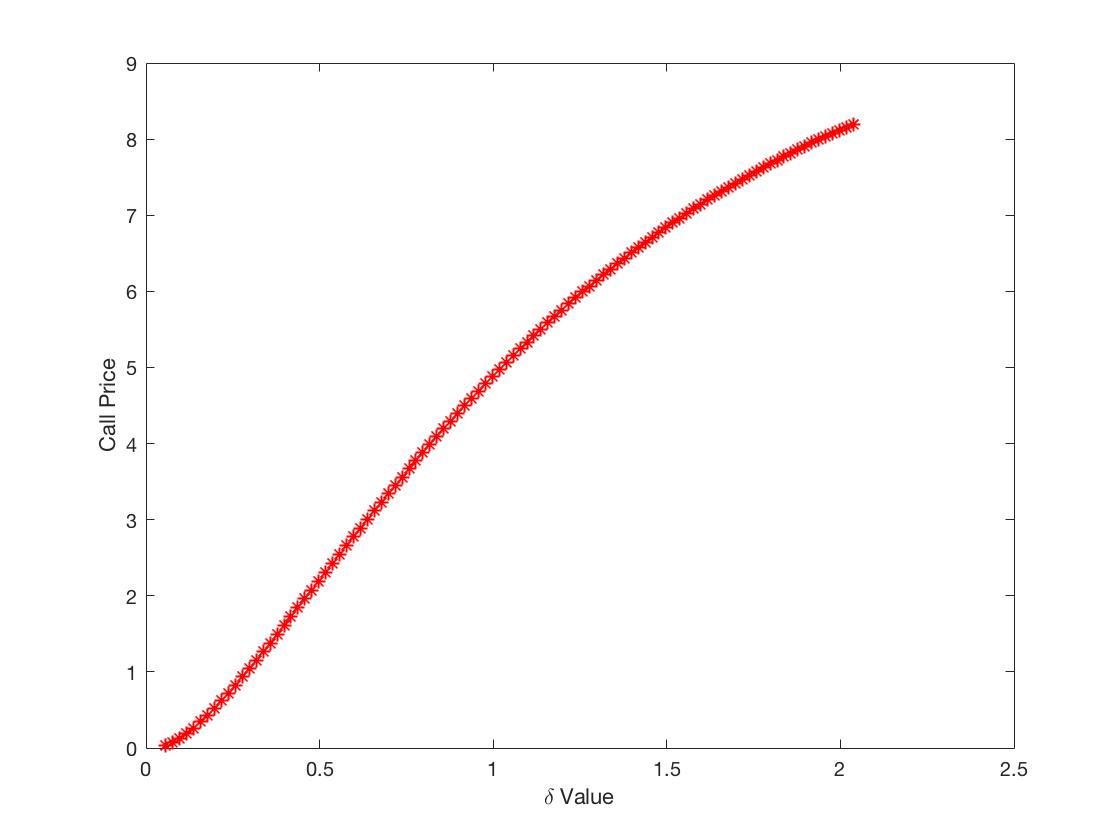}
\label{fig:GHtodelta}}
\subfloat[Subfigure 9 list of figures text][Call price - $\nu$]{
\includegraphics[width=0.3\textwidth]{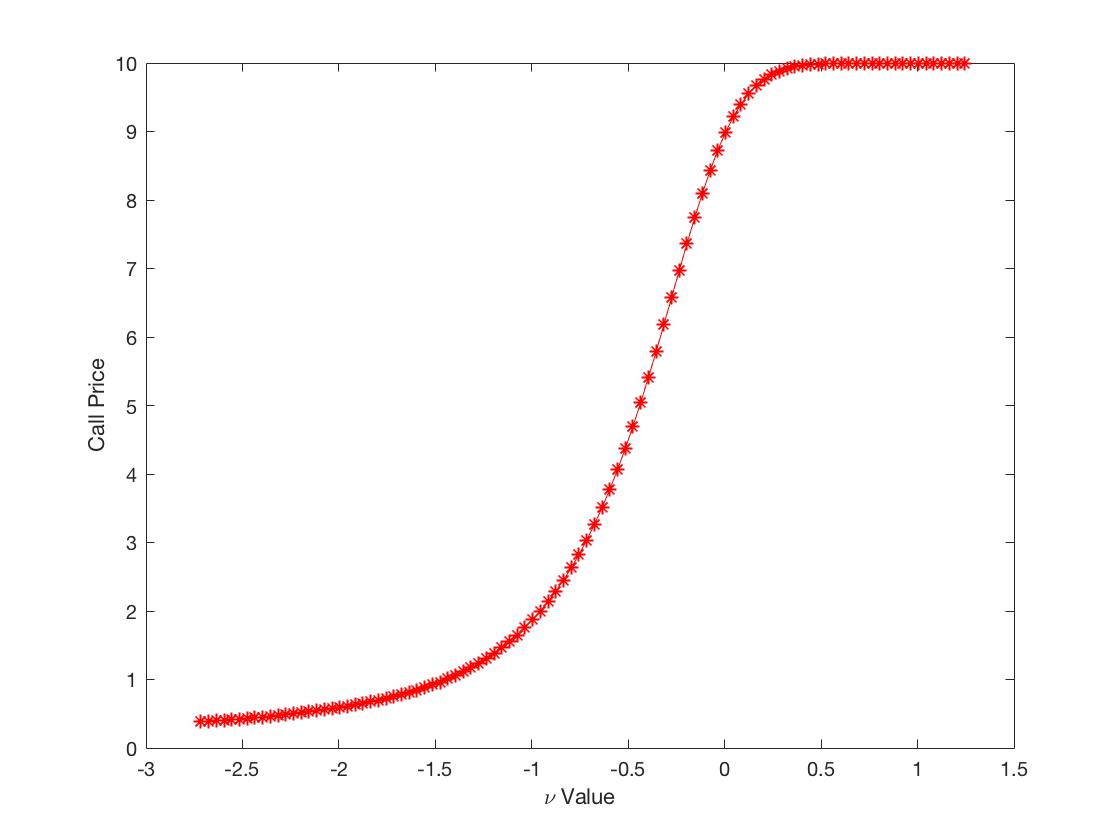}
\label{fig:GHtonu}}

\centering
\caption{GH model parameters' sensitivities.} 

\flushleft

For the basis data,  $n=16$, $S_t=10$, $K=12$, $T=2$, $r=0.05$, $q=0$, $\alpha=3.8288$, $\beta=-3.8286$, $\delta=0.2375$, $\nu=-1.7555$.
The parameter changing setups list as following:
(a) $S_t$ is from 9 to 12, 
(b) $K$ is from 9 to 12, 
(c) $T$ is from 1 to 3, 
(d) $r$ is from 0.01 to 0.11, 
(e) $q$ is from 0 to 0.1, 
(f) $\alpha$ is from $0.8288$ to $5.8288$, 
(g) $\beta$ is from $-3.8286$ to $3.8286$, 
(h) $\delta$ is from $0.0375$ to $2.0375$, 
(i) $\nu$ is from $-2.7555$ to $1.2445$.
In all, the price is increasing with respect to $S_t,T,r,\delta$ and $\nu$, is decreasing with respect to $K$ and $q$. The price is a sort of symmetric with respect to $\alpha$. And there is no tendency about $\beta$.

\label{fig:GHpara}
\end{figure}

For the GH model, we set $\alpha = 3.8288, \beta = -3.8286, \delta = 0.2375, \nu = -1,7555$ from \cite{schoutens2003levy} to understand the sensitivities of the option price with respect to various parameters.
With this data, we compute call prices around the existing parameters by using FFT method with the given characteristic function. Visualizing the trends, we get nine figures listed in Figure (\ref{fig:GHtoS}) to (\ref{fig:GHtonu}).

Figure 1 illustrates the call option prices under the GH model for the underlying asset increases with respect to the underlying asset price $S_t$, the time to maturity $T$, the risk-free rate $r$ and the parameter $\delta$, decreases with respect to the strike price K and the dividend rate $q$. The call option price presents a hump shape with respect to the parameter $\alpha$,
a concave up shape with respect to the parameter $\beta$, a concave up increasing and then concave down increasing to near flat shape with respect to the parameter $\nu$ for the GH model.

\vspace{.15 in}
\textbf{2. Normal Inverse Gaussian (NIG) Distribution}

Setting $\nu = -1/2$, we get the Normal Inverse Gaussian (NIG) distribution from the hyperbolic model in \cite{barndorff1997processes}. The characteristic function of the NIG distribution is given by
$$\phi_{\alpha,\beta,\delta,\mu} 
	=  exp\{\delta \sqrt{\alpha^2-\beta^2}-\delta\sqrt{\alpha^2-(\beta+iu)^2}\}.$$

\begin{figure}[htbp!]
\centering

\subfloat[Subfigure 1 list of figures text][Call prices - $S_t$]{
\includegraphics[width=0.3\textwidth]{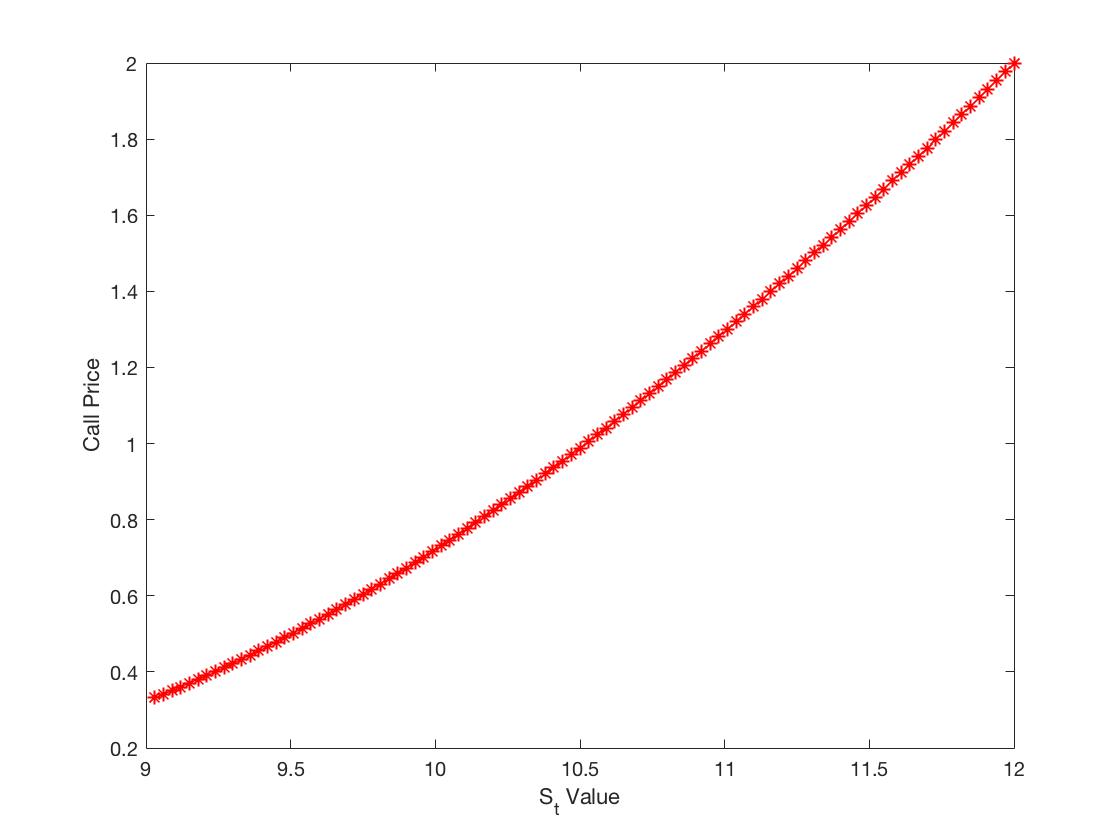}
\label{fig:NIGtoS}}
\subfloat[Subfigure 2 list of figures text][Call prices - K]{
\includegraphics[width=0.3\textwidth]{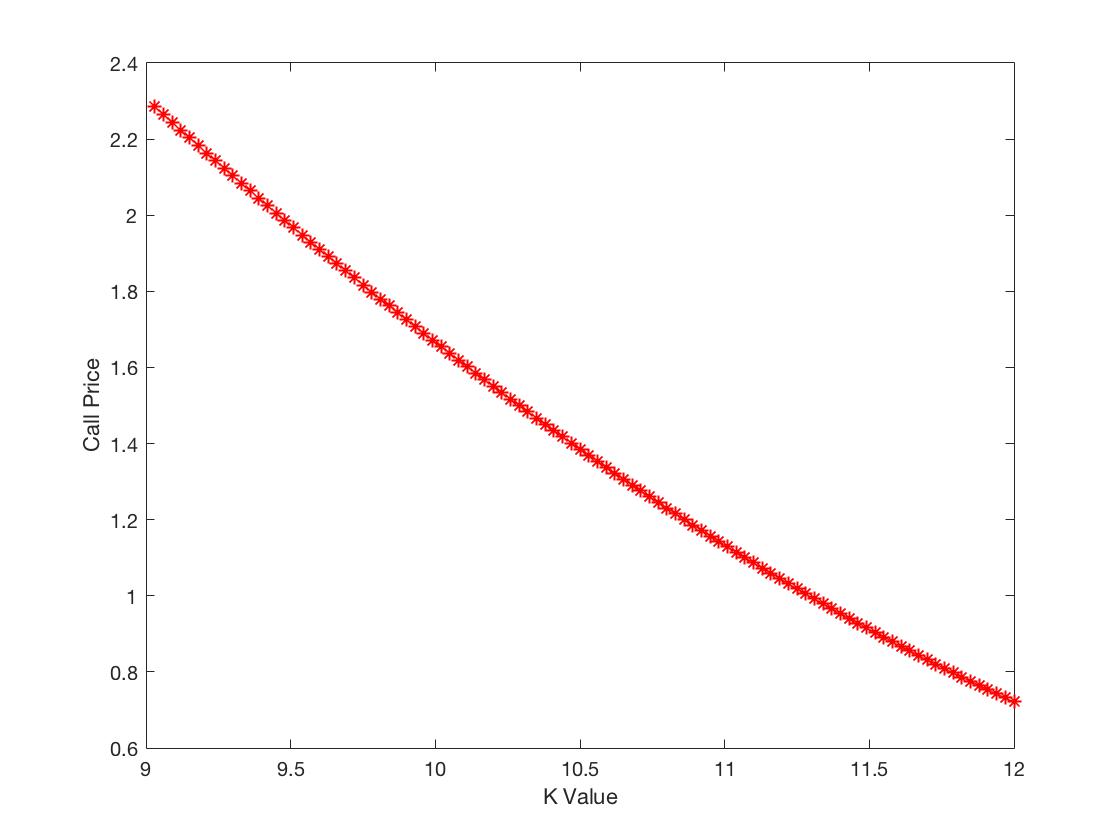}
\label{fig:NIGtoK}}
\subfloat[Subfigure 3 list of figures text][Call price - T]{
\includegraphics[width=0.3\textwidth]{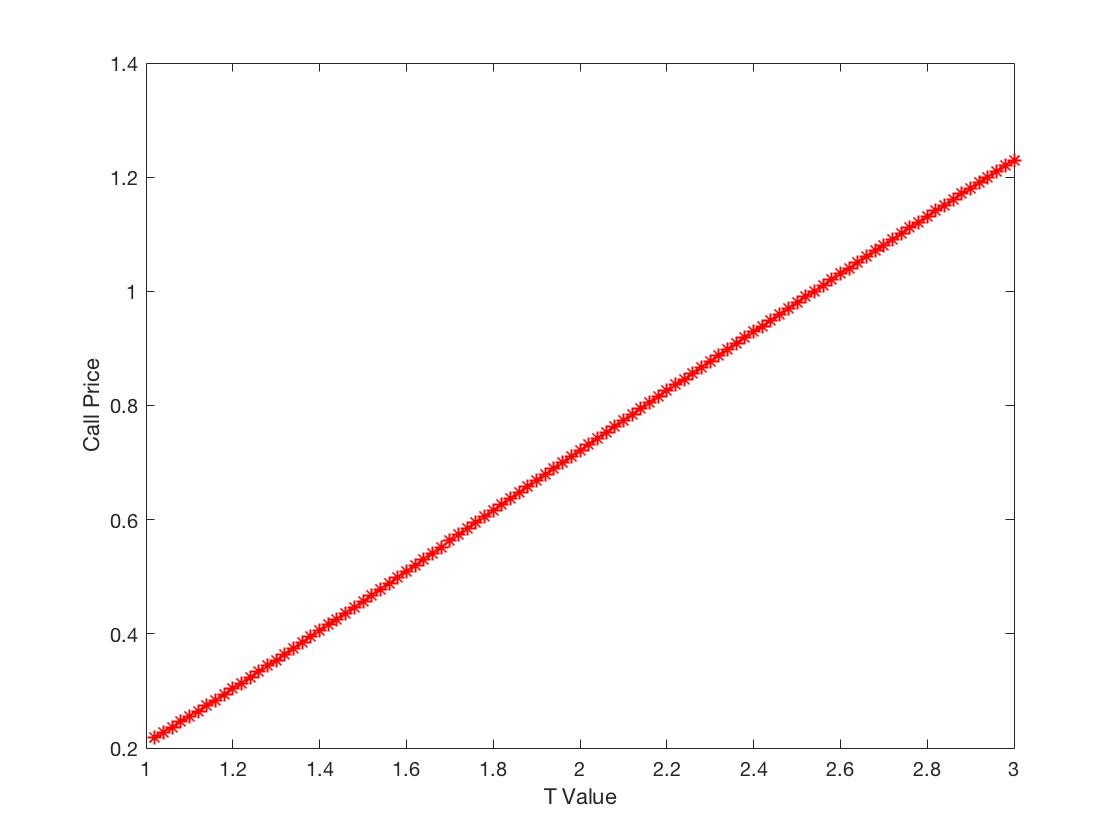}
\label{fig:NIGtoT}}

\subfloat[Subfigure 4 list of figures text][Call price - r]{
\includegraphics[width=0.3\textwidth]{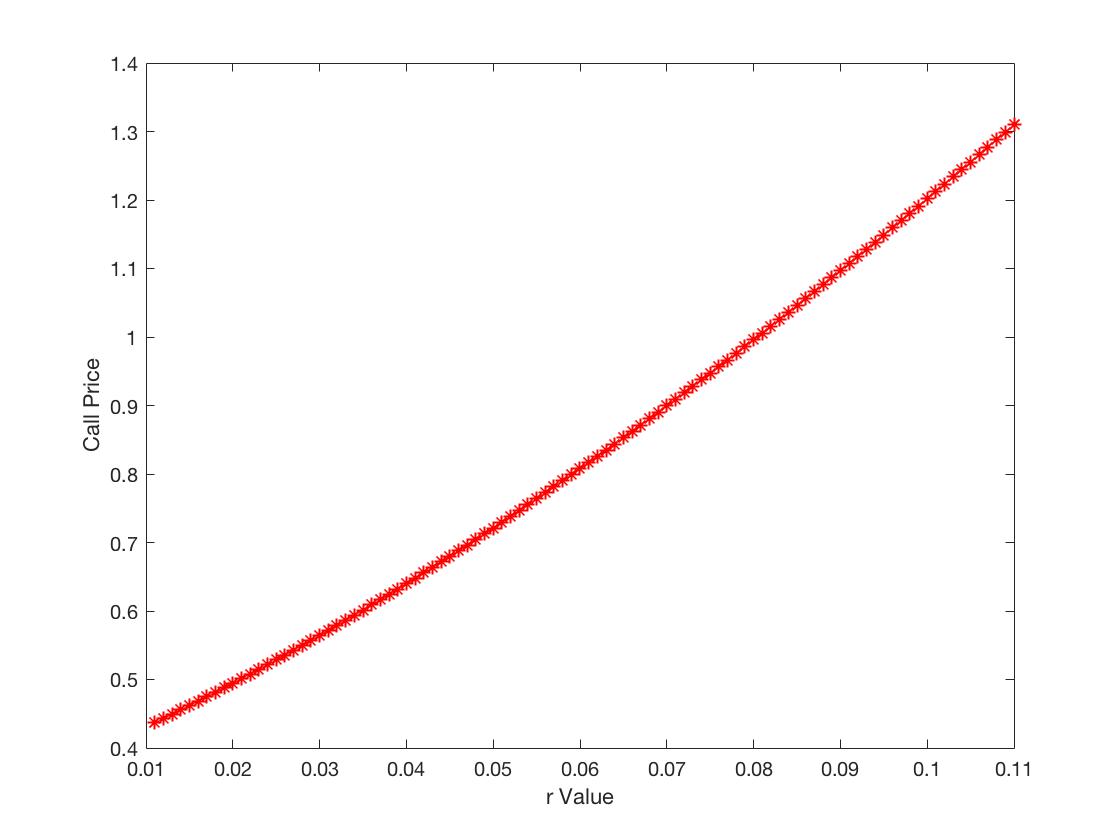}
\label{fig:NIGtor}}
\subfloat[Subfigure 5 list of figures text][Call price - q]{
\includegraphics[width=0.3\textwidth]{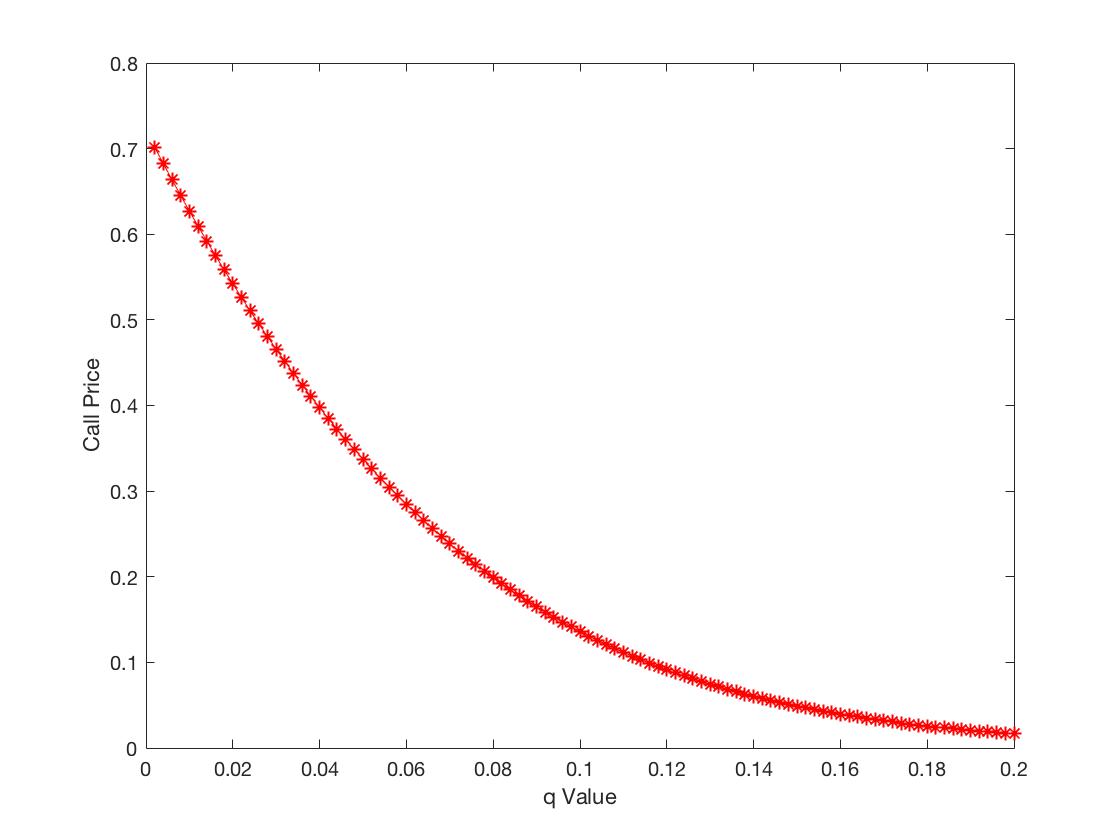}
\label{fig:NIGtoq}}
\subfloat[Subfigure 6 list of figures text][Call price - $\alpha$]{
\includegraphics[width=0.3\textwidth]{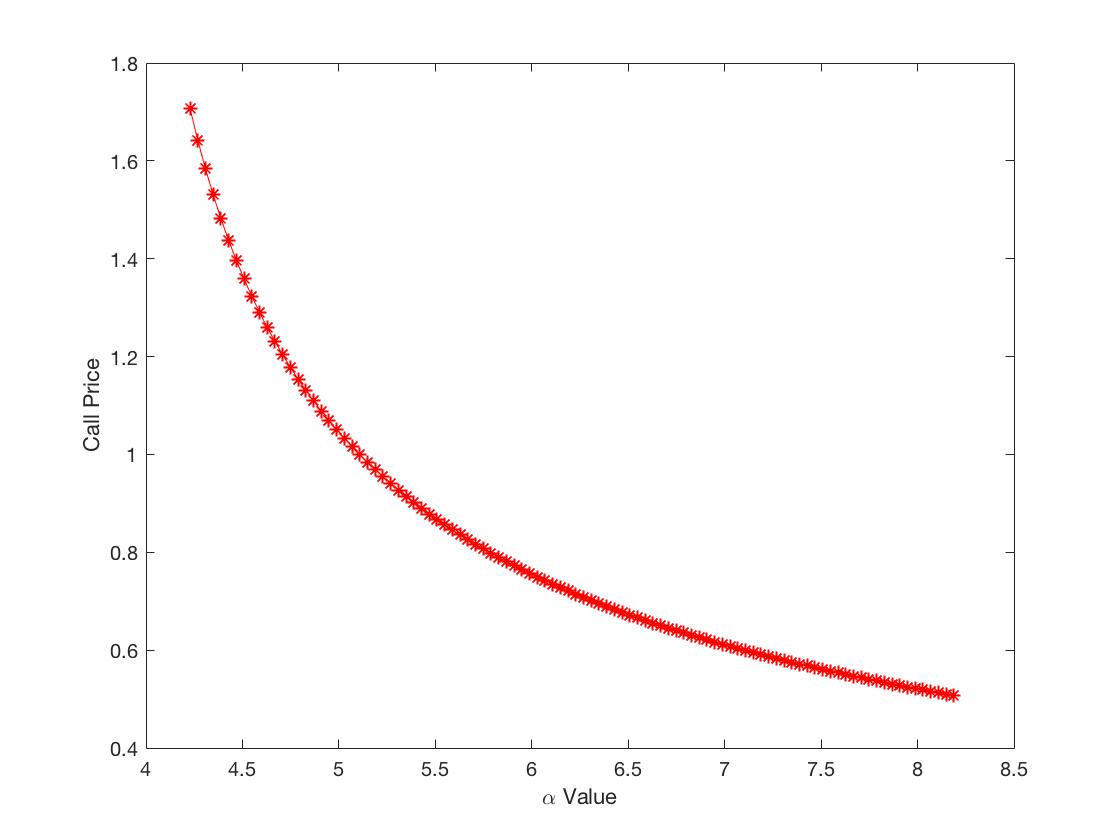}
\label{fig:NIGtoalpha}}

\subfloat[Subfigure 7 list of figures text][Call price - $\beta$]{
\includegraphics[width=0.3\textwidth]{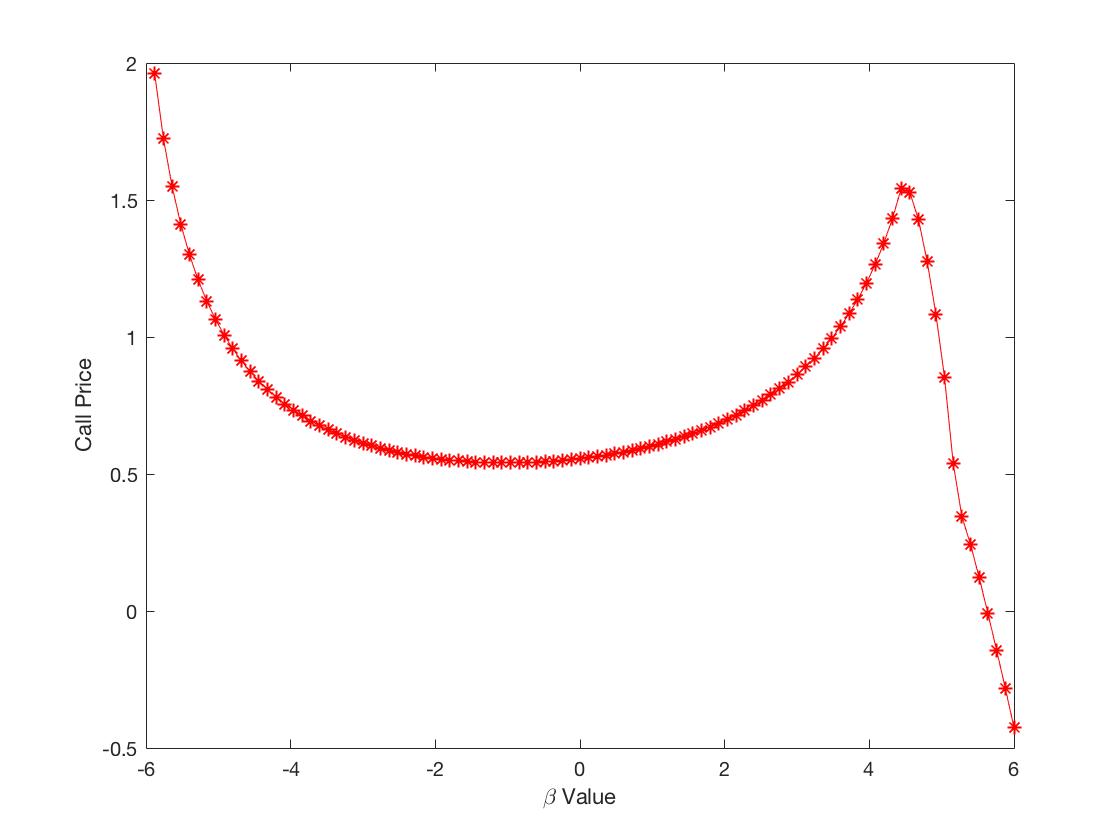}
\label{fig:NIGtobeta}}
\subfloat[Subfigure 8 list of figures text][Call price - $\delta$]{
\includegraphics[width=0.3\textwidth]{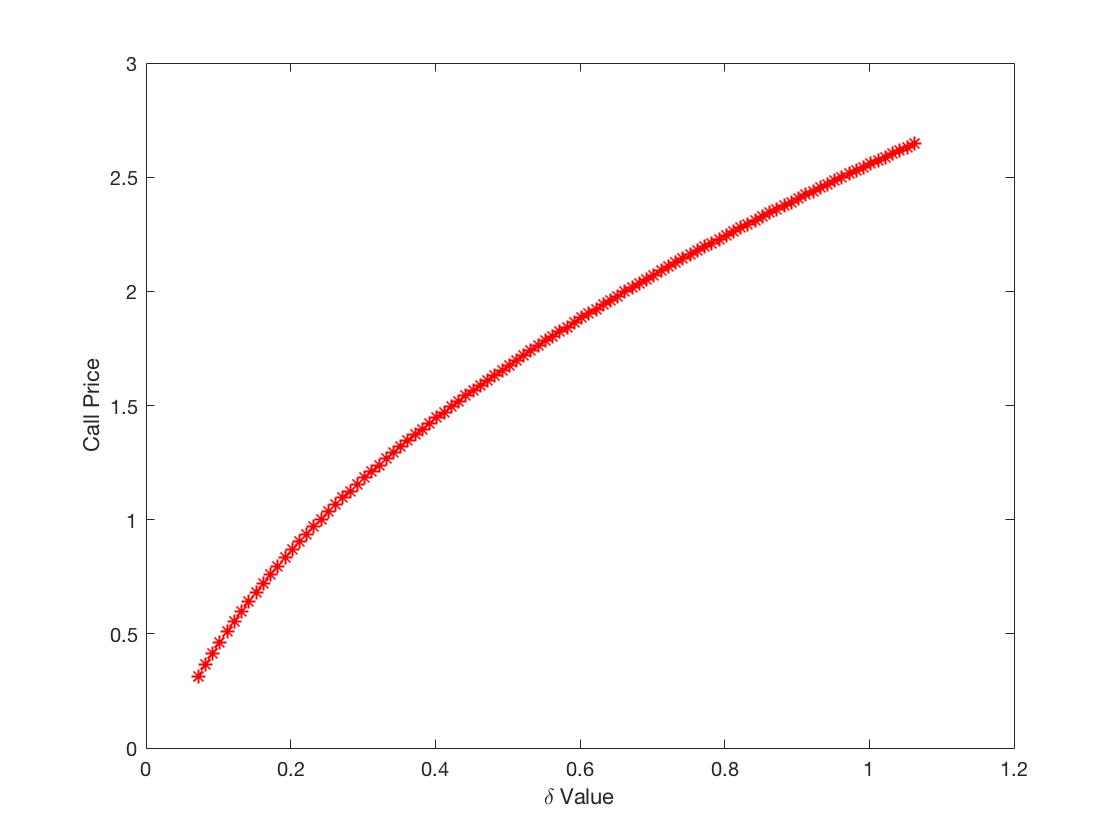}
\label{fig:NIGtodelta}}
\subfloat[Subfigure 9 list of figures text][Call price - $\mu$]{
\includegraphics[width=0.3\textwidth]{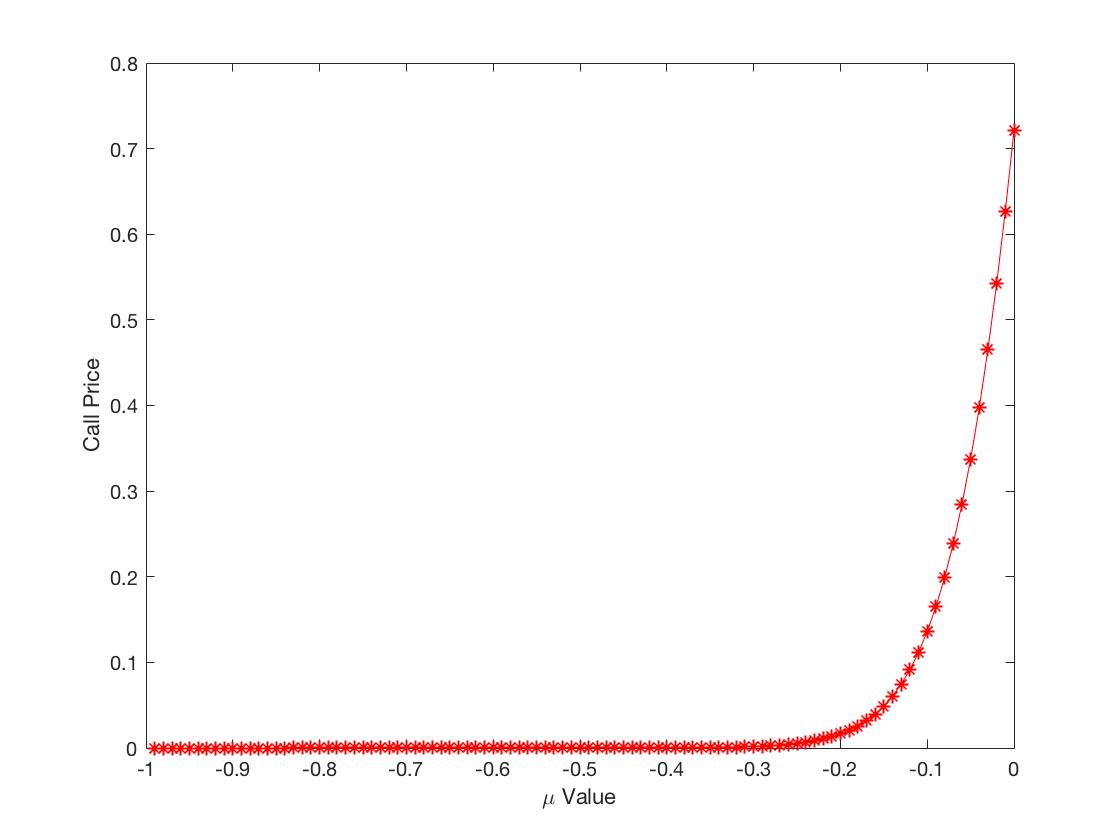}
\label{fig:NIGtomu}}
\centering
\caption{NIG model parameters' sensitivities.} 

\flushleft

For the basis data,  $n=16$, $S_t=10$, $K=12$, $T=2$, $r=0.05$, $q=0$, $\alpha=6.1882$, $\beta=-3.8941$, $\delta=0.1622$, $\mu=0$.
The parameter changing setups list as following:
(a) $S_t$ is from 9 to 12,
(b) $K$ is from 9 to 12, 
(c) $T$ is from 1 to 3, 
(d) $r$ is from 0.01 to 0.11, 
(e) $q$ is from 0 to 0.2, 
(f) $\alpha$ is from $4.1882$ to $8.1882$, 
(g) $\beta$ is from $-6$ to $6$, 
(h) $\delta$ is from $0.0375$ to $2.0375$,
(i) $\mu$ is from $-1$ to $1$.
In all, the price is increasing with respect to $S_t,T,r,\delta$ and $\mu$, is decreasing with respect to $K$, $q$ and $\alpha$. And there is no tendency about $\beta$.
\label{fig:NIGpara}
\end{figure}

For the NIG model, we set $\alpha= 6.1882, \beta = -3.8941, \delta = 0.1622$  from \cite{schoutens2003levy} to understand the sensitivities of the option price with respect to various parameters. We set  $\mu = 0$ as the initial value for NIG model since it is the drifting term of process.
Therefore, we obtain Figure (\ref{fig:NIGpara}) by FFT method estimation for existing characteristic function of the NIG model.

For the NIG model, Figure 2 shows that the option price is increasing with respect to $S_t,T,r,\delta$ and $\mu$, is decreasing with respect to $K$, $q$ and $\alpha$. And there is no trend about $\beta$, where it exhibits the concave up shape connecting with a deep down. For the parameter $\beta$, the behavior is undetermined since the NIG is a special case of the GH model.
The option price is increasing concave down way for the parameter $\delta$, and concave up way for the parameter $\mu$. 

\vspace{.15 in}

\textbf{3. The Carr-Geman-Madan-Yor (CGMY) Class of Distribution}

The CGMY distribution class is defined by \cite{carr2002fine} with the L\'evy density 
$$ K_{CGMY} = \frac{C}{|x|^{1+Y}}exp\{\frac{G-M}{2}x-\frac{G+M}{x}|x|\}.$$
The characteristic function of the CGMY distribution is 
$$\phi_{CGMY} = exp \{ C\Gamma (-Y)[(M-iu)^Y-M^Y+(G+iu)^Y-G^Y] \}.$$
The CGMY model is also the class of tempered stable processes  with time
changed Brownian motion, it has a better model freedom than those with Brownian
subordination. The main impact on option prices is due to large jumps, and the CGMY model shows its flexibility among those tempered stable processes.

\begin{figure}[htbp!]
\centering
\subfloat[Subfigure 1 list of figures text][Call prices - $S_t$]{
\includegraphics[width=0.3\textwidth]{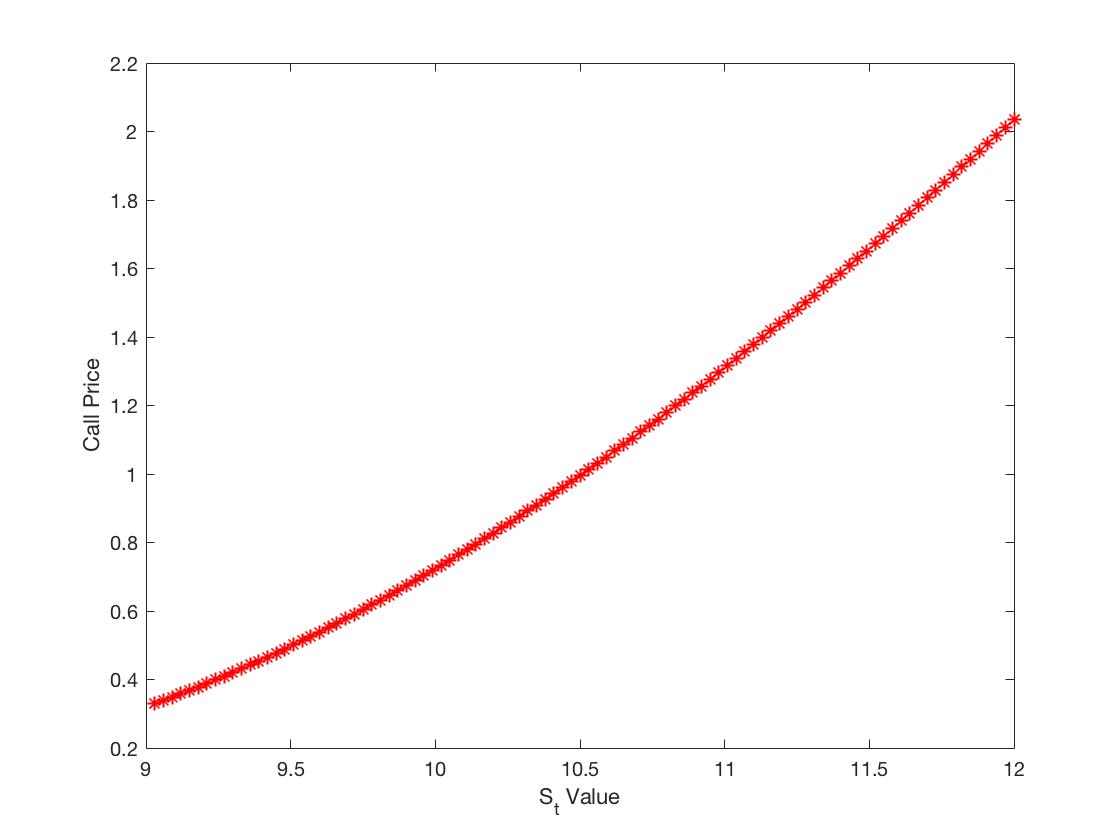}
\label{fig:CGMYtoS}}
\subfloat[Subfigure 2 list of figures text][Call prices - K]{
\includegraphics[width=0.3\textwidth]{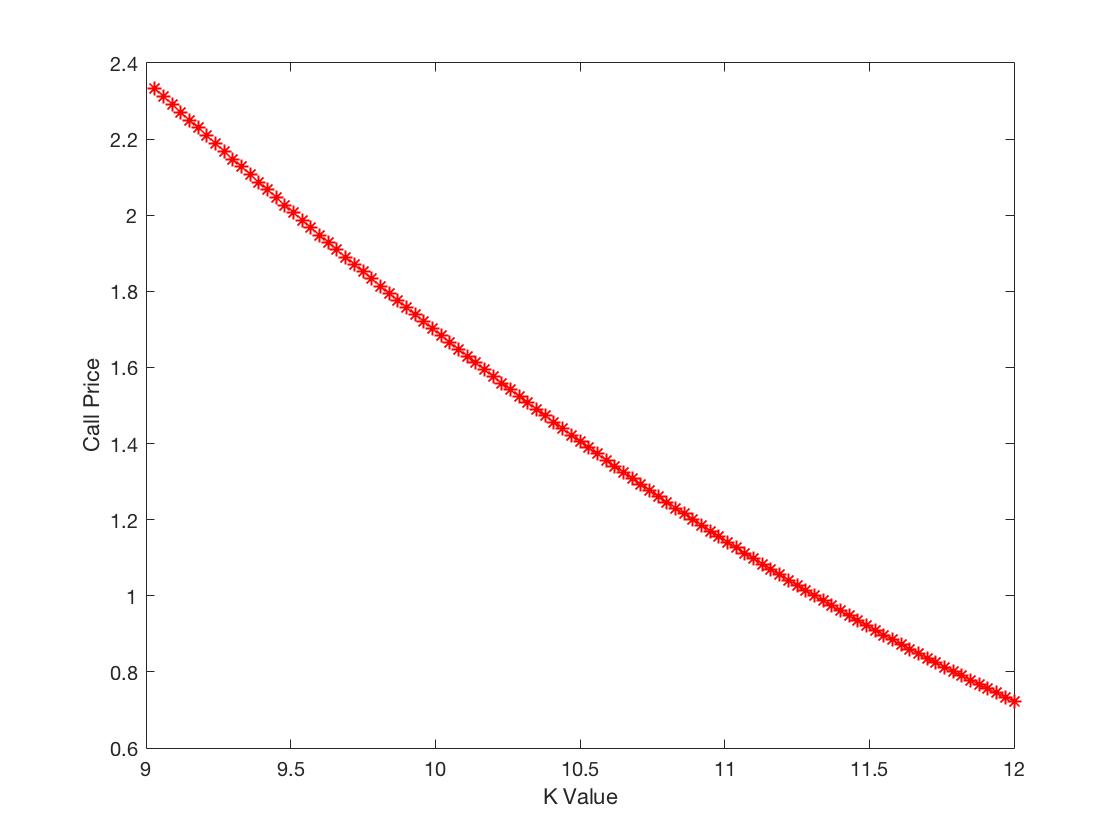}
\label{fig:CGMYtoK}}
\subfloat[Subfigure 3 list of figures text][Call price - T]{
\includegraphics[width=0.3\textwidth]{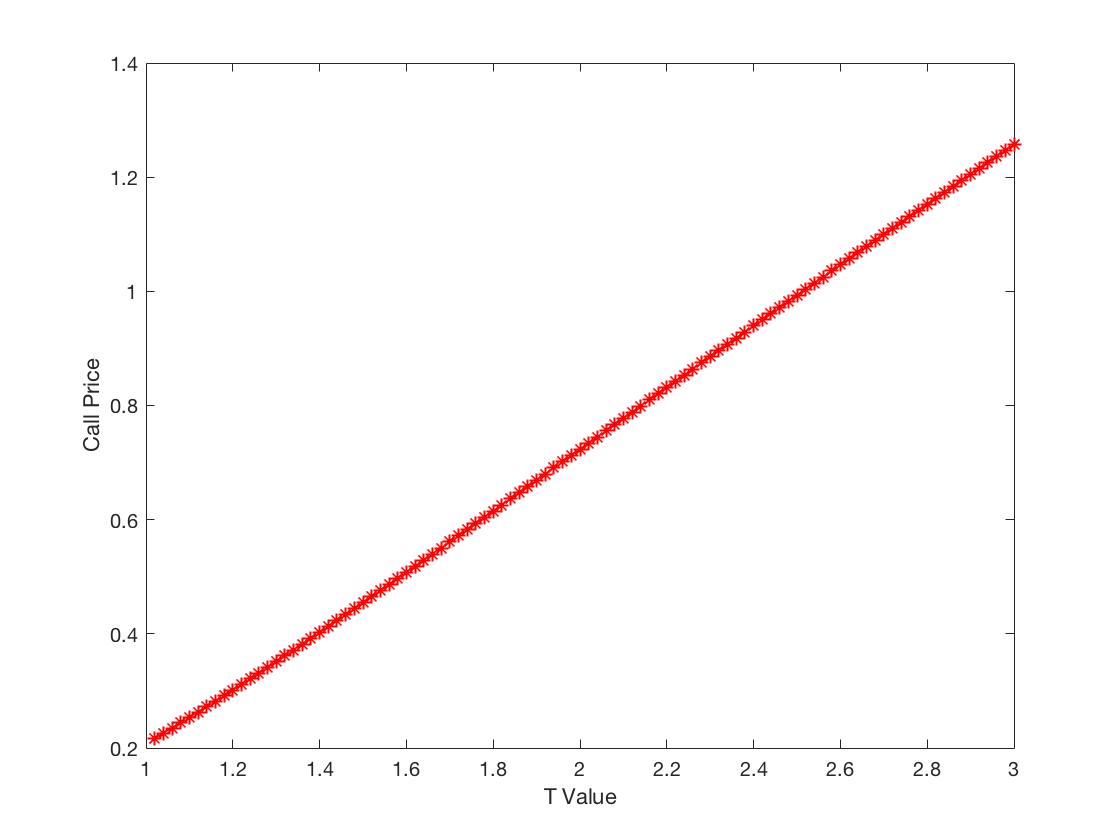}
\label{fig:CGMYtoT}}

\subfloat[Subfigure 4 list of figures text][Call price - r]{
\includegraphics[width=0.3\textwidth]{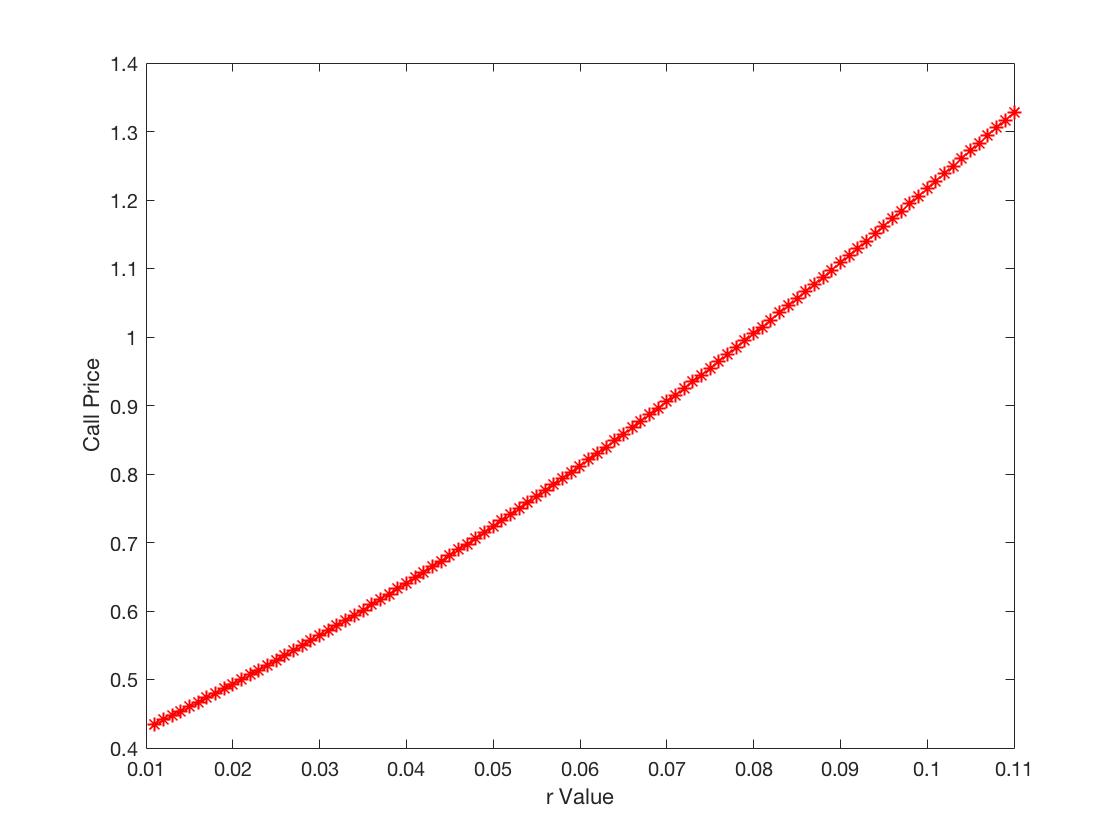}
\label{fig:CGMYtor}}
\subfloat[Subfigure 5 list of figures text][Call price - q]{
\includegraphics[width=0.3\textwidth]{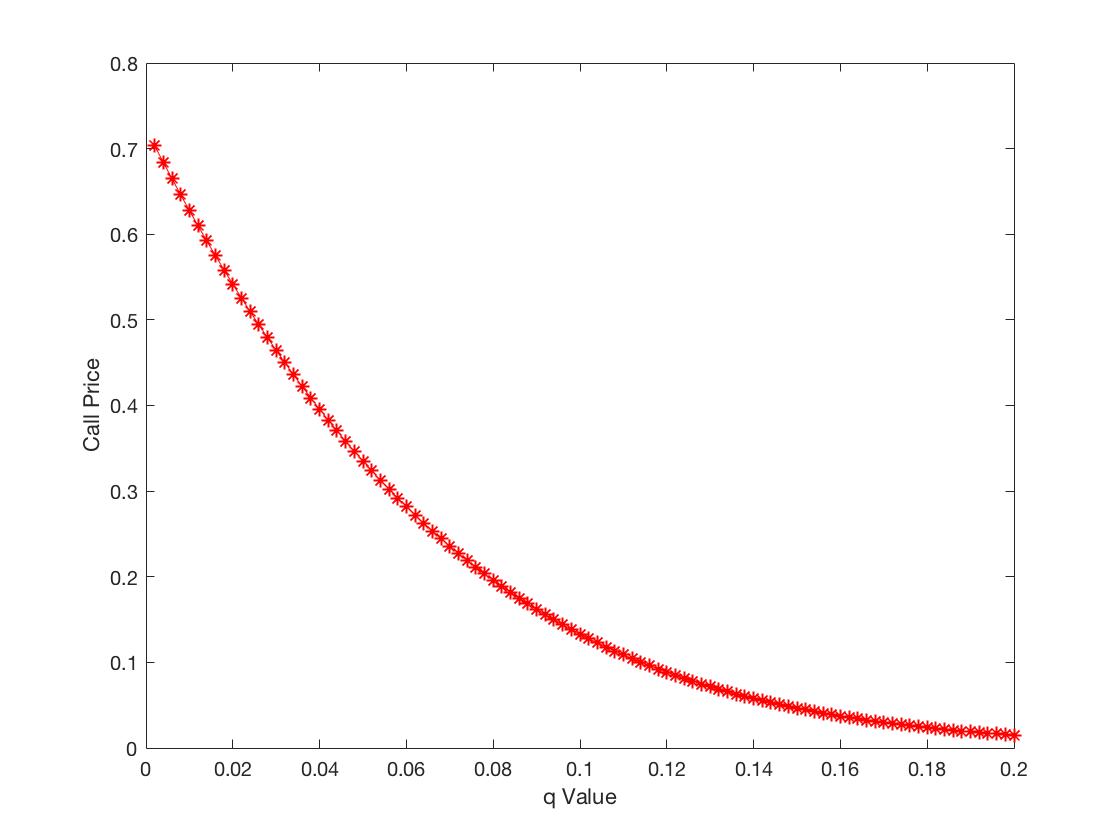}
\label{fig:CGMYtoq}}
\subfloat[Subfigure 6 list of figures text][Call price - C]{
\includegraphics[width=0.3\textwidth]{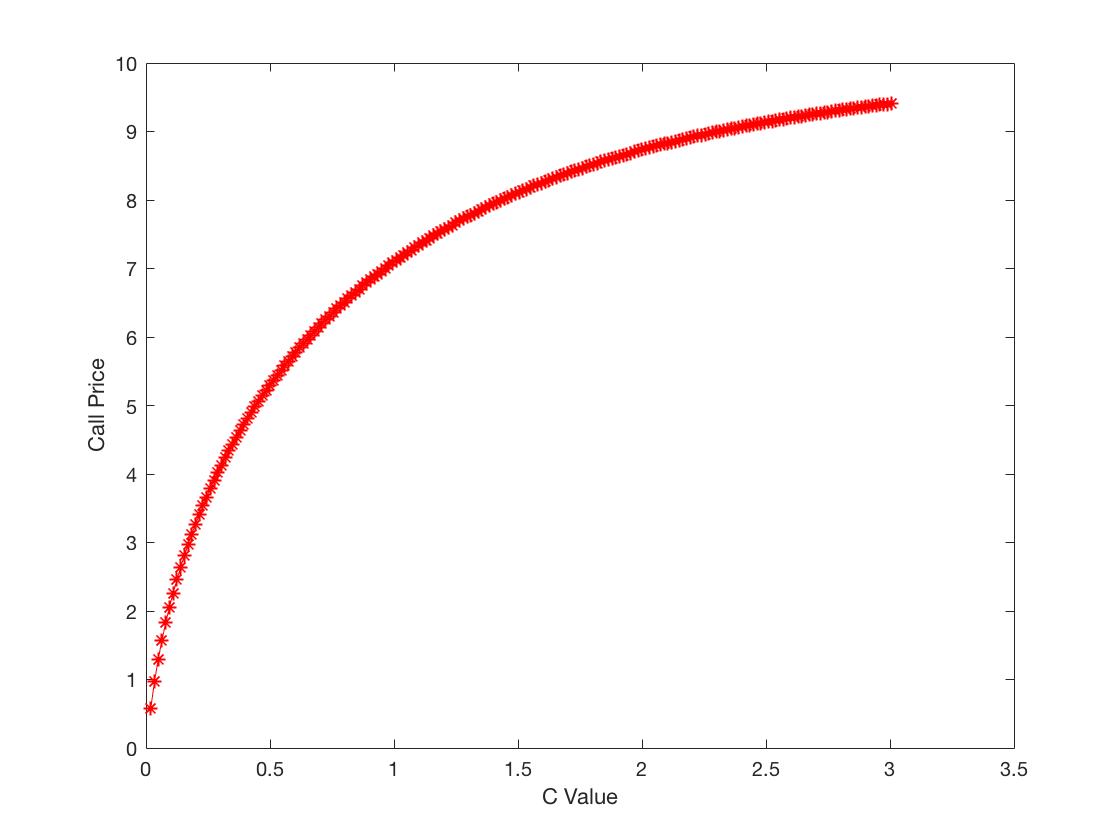}
\label{fig:CGMYtoC}}

\subfloat[Subfigure 7 list of figures text][Call price - G]{
\includegraphics[width=0.3\textwidth]{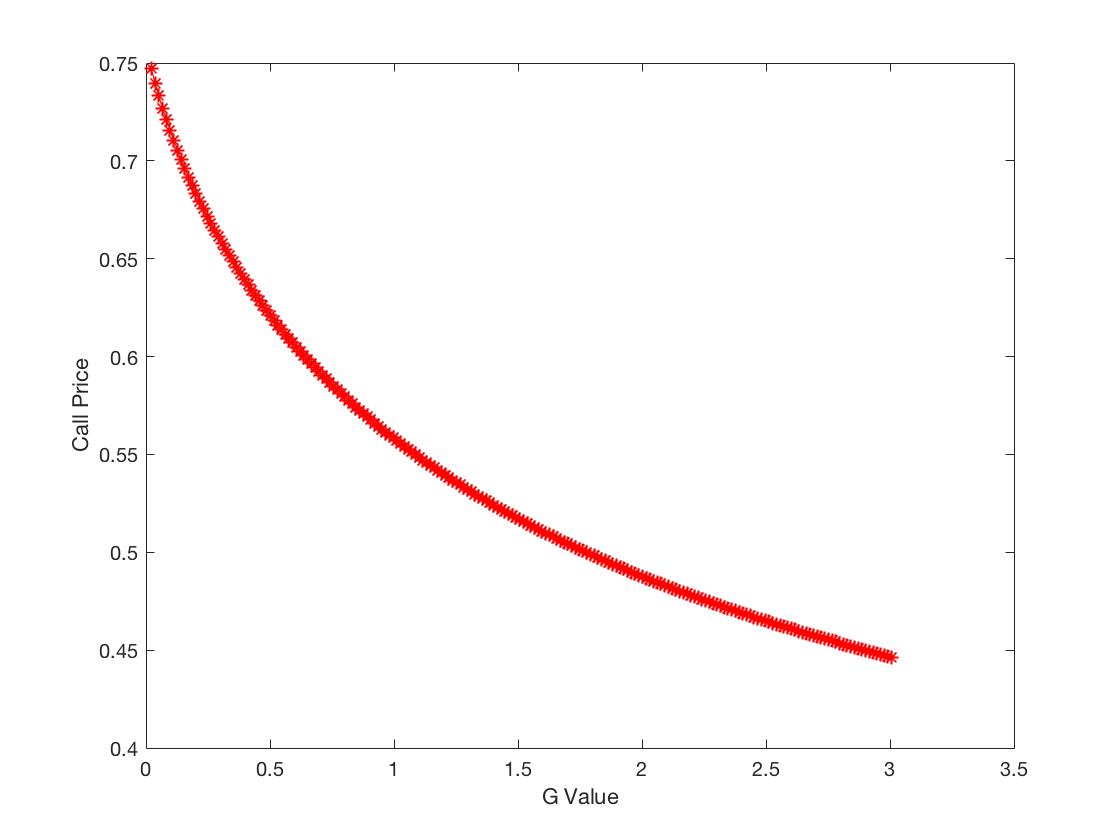}
\label{fig:CGMYtoG}}
\subfloat[Subfigure 8 list of figures text][Call price - M]{
\includegraphics[width=0.3\textwidth]{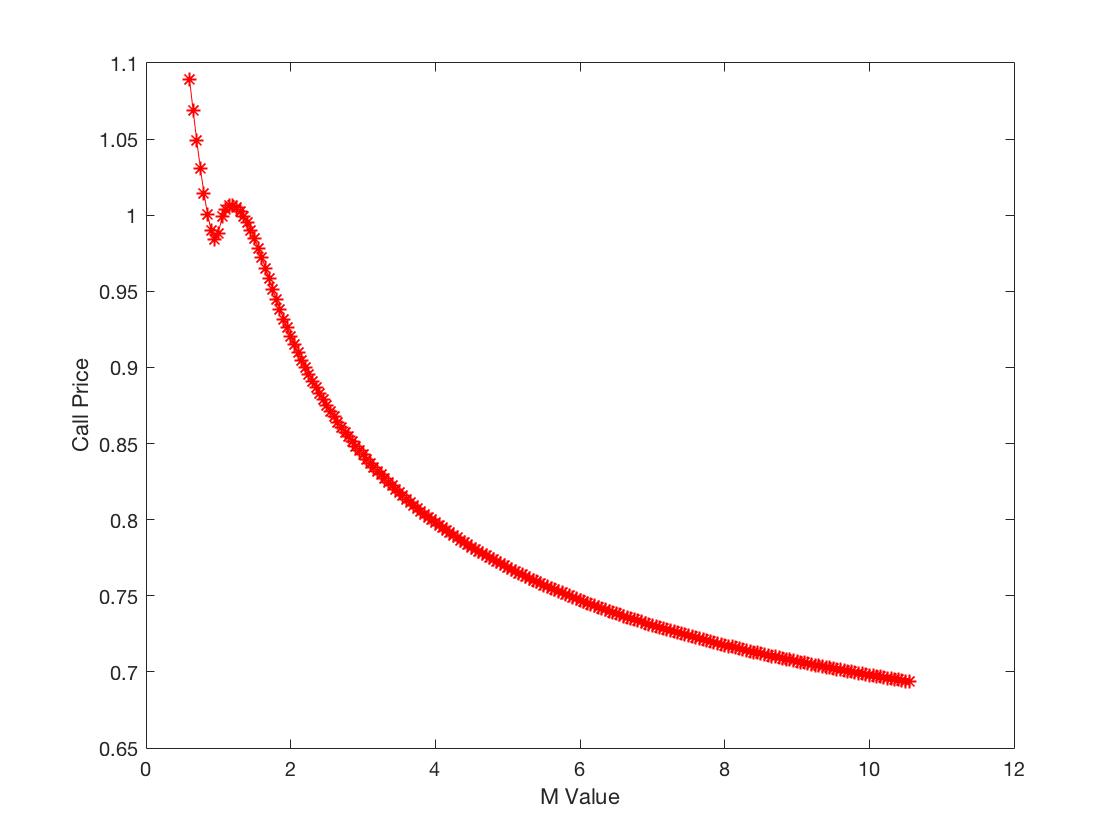}
\label{fig:CGMYtoM}}
\subfloat[Subfigure 9 list of figures text][Call price - Y]{
\includegraphics[width=0.3\textwidth]{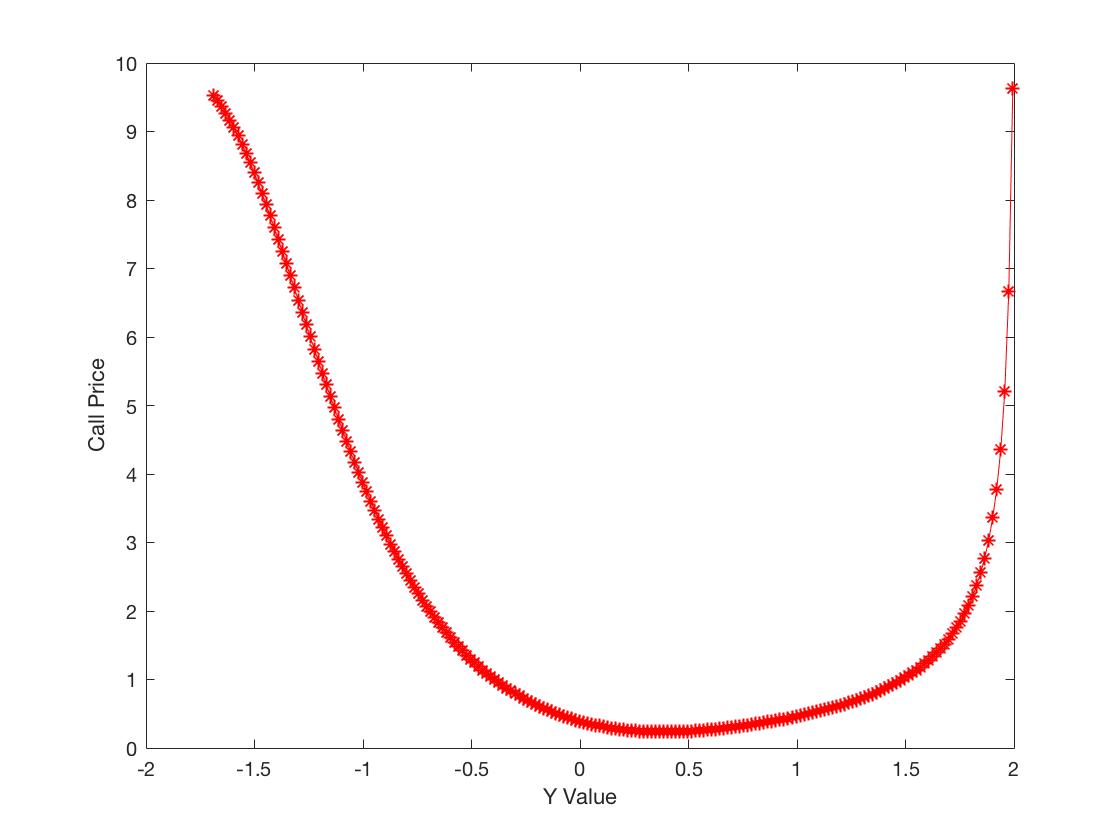}
\label{fig:CGMYtoY}}

\caption{CGMY model parameters' sensitivities.} 

\flushleft
For the basis data,  $n=16$, $S_t=10$, $K=12$, $T=2$, $r=0.05$, $q=0$, $C=0.0244$, $G=0.0765$, $M=7.5515$, $Y=1.2945$.
The parameter changing setups list as following:
(a) $S_t$ is from 9 to 12, 
(b) $K$ is from 9 to 12,  
(c) $T$ is from 1 to 3,  
(d) $r$ is from 0.01 to 0.11, 
(e) $q$ is from 0 to 0.2, 
(f) $C$ is from 0.0044 to 3.0044,  
(g) $G$ is from 0.0065 to 3.0065,  
(h) $M$ is from 0.5515 to 10.5515,  
(i) $Y$ is from -1.7055 to 1.9945. 
In all, the price is increasing with respect to $S_t, T, r$ and $C$, is decreasing with respect to $K$, $q$ and $G$. And there are no uniform tendencies about $M$ and $Y$.
\label{fig:CGMYpara}
\end{figure}

For the CGMY model, we set $C=0.0244,G= 0.0765,M= 7.5515,Y= 1.2945$ from \cite{schoutens2003levy}. The FFT method for estimating call option prices gives Figures (\ref{fig:CGMYpara}) about parameter sensitivities.

For the CGMY model over our data range, the option price is increasing with respect to $S_t, T, r$, and increasing concave down way in $C$, is decreasing with respect to $K$, and decreasing concave up way in $q$ and $G$. And the tendency about $M$ is increasing with a small hump and then starts to decrease. The option price has the U-shape with respect to the parameter Y in the CGMY model.

\section{Empirical Analysis on S\&P 500 Index Option}
In this section, we use the sum of squared pricing error (SSE) method to calibrate the model parameters by optimizing the least squared summation as \cite{bakshi1997empirical} did for those five models. We focus on the BS, SV, Non-iid, L\'evy processes (GH, NIG, CGMY) models to compare the relative performance on these variations. First we shall use the data to examine the direction and the extent of biases associated with the BS, to back out a BS implied volatility from each option price in the sample. 

 The sum of squared dollar pricing errors (SSE) is obtained by following. Collect $N$ option prices on S\&P 500 option taken from the same period. For each $n=1,...,N$, setting $\tau_n$ and $K_n$ be the time-to-maturity and the strike prices of the $n$-th option respectively. Let $\hat{C}_n(t,\tau_n,K_n)$ be the observed market option price and $C_n(t,\tau_n,K_n)$ the model price determined by equation (\ref{HestonC}) for BS, SV and equation (\ref{Li15}) for non-iid and equation (\ref{PIDE}) for the three typical L\'evy jump models. The difference between $\hat{C}_n$ and $C_n$ is a function of $V(t)$ and parameters.
Define 
$\epsilon_n = \hat{C}_n(t,\tau_n,K_n)-C_n(t,\tau_n,K_n)$.
Then SSE is defined by 
\begin{equation}\label{sse}
SSE(t) = min \sum^N_{n=1}|\frac{\epsilon_n}{BSVega}|^2.
\end{equation}
where $BSVega$ is the Black-Scholes sensitivity of the option price with respect to the market implied volatility $V_t$:
$$BSVega = S e^{(-q\tau_n)}n(d_n)\sqrt{\tau_n}, \ \ 
 d_n=\frac{ln(S/K_n)+(r-q+V^2_n/2)\tau_n}{V^2_n\sqrt{\tau_n}}, $$
and $n(x)=e^{-x^2/2}/\sqrt{2\pi}$ is the standard normal density (see \cite{lewisoption} or \cite{christoffersen2009shape}). Note that our $SSE(t)$
is normalized by the Vega refer to \cite{christoffersen2009shape}, where \cite{bakshi1997empirical} simply use the
squared dollar pricing errors without the $BSVega$ term.

Considering the S\&P 500 index option during September 2012 to August 2013, the previous half year's data, from September 4th 2012 to February 28th 2013, is treated as in-sample and the latter period, from March 1st 2013 to August 30th 2013, as out-of-sample. For the spot volatility which is conditional on no jump from the BS model, the structural parameters and the jump-related parameters as well as parameters of three typical L\'evy models are required to be
estimated. One may apply econometric tools to obtain these required
estimations. However, in order to be consistent with the empirical results
along \cite{bakshi1997empirical}, we follow their method to reduce data
requirement dramatically (comparing to other methods) and also improve the
performance significantly.

\subsection{In-sample Pricing Fit}

\vspace{.15 in}
\textbf{1. Comparison of BS model and Heston Model }

For the in-sample data, we need to find the optimal parameters in the close-form Heston option price formula (\ref{HestonC}) firstly to fix those continuous model parameters (no jumps). We minimize (\ref{sse}) by applying an implied  parameter procedure to implement candidate models to best estimate each model parameters. 

The parameters are estimated by $fmincon$ function in Matlab to find the group of parameters which make the SSE value in (\ref{sse}) to be the minimum. 
The $fmincon$ is an important gradient-based algorithm of Matlab to find a constrained minimum for a multivariable function for nonlinear optimization starts at an initial estimate. Based on previous results in 
\cite{bakshi1997empirical} and \cite{rouah2013heston}, we can have a rough idea to restrict each parameter in its searching region. In our experiments, the searching initial and range of parameters are listing below.
The parameter $\kappa$ starts at 2 in the range $[0,20]$, the parameter 
$\theta$ starts with $0.05$ in the range $[0,2]$, $\sigma$ with initial parameter 1.3 lies in $[0,2]$, the prarmeter $\rho$ starts at 0.8 in the range $[-1,1]$. Similarly we have that $\lambda$ starts at 0.05 in $[0,2]$, $\mu$ starts at -0.1 in $[-1,1]$ and $\sigma_j$ starts at 0.1 in $[0,2]$. The volatility $V_t$ starts at 0.5 in $[0,1]$ from earlier empirical estimates \cite{bakshi1997empirical}, \cite{rouah2013heston} etc.

For the in-sample data, there are total $27,363$ option prices for S\&P 500 index. We handle two panels of different upper bound as 100 and 500 for the integral in the equation (\ref{HestonPi}) due to different computational consumption. The parameters are listed in Table \ref{Hestonpara}.

The results tell us the two panels have the same order of magnitude of SSE values although the panel B has smaller numbers. We prefer to use results of integral upper bound as 100 in panel A to compare with L\'evy processes calibration later since it has lower computational consumption.
For different time periods, we can see those parameters varying dramatically 
as our Table 3 for the S\&P 500 index option on September 4, 2012-- Feburary  28, 2013 and Table III for the S\&P 500 index option on June 1988--May 1991 of
\cite{bakshi1997empirical}.

For every panel, we deal with the diffusion volatility in two cases. One case is to use the implied volatility  given by the market noted as $imp.V_t^2$. Another case is to estimate that parameter as a fixed value in the optimal searching. 

\begin{table}[]
\small
\raggedright
\caption{Heston Model Parameters Optimal Estimation for SV and SVJ Models.}
The structural parameters of SV and SVJ  are estimated by minimizing the sum of squared pricing errors between the market price and the model-determined price for options in each day in the sample period.  The SSE in (\ref{sse}) with weighted BSVega is estimated in Panel A with the upper bound 100 and Panel B with the upper bound 500. For each model in each panel, there are two ways to choose the volatility parameter, first column is estimated by using the implied volatility and the second column by estimating the volatility parameter. The results show that the two panels have the same order of magnitude of SSE values although  Panel B has smaller numbers.
\label{Hestonpara}
\begin{tabular}{lllllllllllll}
\\
 \hline
 &  & \multicolumn{5}{l}{Panel A (upbound=100)} &  & \multicolumn{5}{l}{Panel B (upbound=500)} \\ \cline{3-7} \cline{9-13} 
 &  & SV &  &  & SVJ &  &  & SV &  &  & SVJ &  \\ \cline{3-4} \cline{6-7} \cline{9-10} \cline{12-13} 
$\kappa$ &  & 20.0000 & 12.8826 &  & 19.9994 & 13.2477 &  & 20.0000 & 10.329 &  & 20.0000 & 11.493 \\
$\theta$ &  & 0.0112 & 0.0125 &  & 0.0109 & 0.0131 &  & 0.0336 & 0.0112 &  & 0.0319 & 0.0108 \\
$\sigma$ &  & 0.7329 & 0.1399 &  & 0.7244 & 0.1551 &  & 2.2288 & 0.3169 &  & 2.1773 & 0.2422 \\
$\rho$ &  & 0.1954 & 0.7938 &  & 0.1921 & 0.6891 &  & 0.0054 & 0.5992 &  & -0.0025 & 0.7659 \\
$V_t$ &  & $imp.V_t^2$ & 0.0089 &  & $imp.V_t^2$ & 0.0090 &  & $imp.V_t^2$ & 0.0076 &  & $imp.V_t^2$ & 0.0077 \\
$\lambda$ &  &  &  &  & 0.5207 & 0.0530 &  &  &  &  & 0.00001 & 0.00001 \\
$\mu_j$ &  &  &  &  & 0.0001 & -0.0001 &  &  &  &  & -0.0592 & -0.3245 \\
$\sigma_j$ &  &  &  &  & 0.0002 & 0.0009 &  &  &  &  & 1.9679 & 0.00009 \\
SSE &  & 5.1238 & 4.1293 &  & 5.1045 & 4.0942 &  & 3.7231 & 3.5773 &  & 3.7668 & 3.5591\\ \hline
\end{tabular}
\end{table}

Table \ref{Hestonpara} illustrates that the SSE results, are equal to
 5.1238 for using the implied volatility in the first case, and equal to 4.1293 for the second case of estimating volatility value of SV model in panel A. 
 We see that the second case of estimating volatility value gets the lower consequence in both panel A and B. And the SSE value drops a bit from 5.1238 of SV model to 5.1045 of SVJ model for the first volatility case in panel A due to the jump, and from 4.1293 of SV model to 4.0942 of SVJ mode for the second estimating volatility parameter. Generally, SVJ model has the lower SSE values except the findings in panel B, which has 3.7231 of SV model improved to 3.7668 of SVJ model in the first volatility case of panel B, from 3.5773 of SV model to 3.5591 of SVJ model for the second estimating volatility parameter. This is a clue for the idea that the Poison jumps part in Heston model contributes less misspecification on option prices.

The comparison with the BS model is listed in Table \ref{comparisonin}.

\begin{table}[]
\small
\raggedright
\caption{Comparison with the BS and the Heston Model for In-sample Data}
Define averages of error (AE) between the Heston model price $C_{Heston}$ and the market price $C_{market}$ as $|C_{Heston}-C_{market}|$ and averages of relative error (ARE) as $|\frac{C_{Heston}-C_{market}}{C_{market}} |$. The AE and ARE values are remarkably dropping from BS model to Heston model for all cases. 
For both implied volatility and estimating volatility parameter cases, the SSE, AE and ARE are improved from the BS model to the Heston model, and from the SV model to the SVJ model for both cases in each corresponding panel, except in Panel B the AE and ARE of the SV model for the implied volatility case are smaller than those of SVJ and the ARE of the SV model for the estimating volatility case is smaller than that of the SVJ model. \\
\label{comparisonin}
\resizebox{\textwidth}{!}{%
\begin{tabular}{llllllllllllll}
\\
\hline
 &  &  & \multicolumn{5}{l}{Panel A (upbound=100)} &  & \multicolumn{5}{l}{Panel B (upbound=500)} \\ \cline{4-8} \cline{10-14} 
 & BS &  & SV &  &  & SVJ &  &  & SV &  &  & SVJ &  \\ \cline{4-5} \cline{7-8} \cline{10-11} \cline{13-14} 
 &  &  & $imp.V_t^2$ & estimate $V_t$ &  & $imp.V_t^2$ & estimate $V_t$ &  & $imp.V_t^2$ & estimate $V_t$ &  & $imp.V_t^2$ & estimate $V_t$ \\ \cline{2-2} \cline{4-5} \cline{7-8} \cline{10-11} \cline{13-14} 
SSE & 96.3520 &  & 5.1238 & 4.1293 &  & 5.1045 & 4.0942 &  & 3.7231 & 3.5773 &  & 3.7668 & 3.5591 \\
AE & 15.8135 &  & 1.9272 & 1.8763 &  & 1.8982 & 1.8305 &  & 1.6279 & 1.6004 &  & 1.6529 & 1.5676 \\
ARE & 4.4620 &  & 0.6035 & 0.5400 &  & 0.5770 & 0.5082 &  & 0.4692 & 0.4123 &  & 0.5139 & 0.4183\\ \hline
\end{tabular}%
}
\end{table}

The Heston model has the sum of squared error dramatically less than BS model by comparing the consequence in Table \ref{comparisonin} for both implied volatility and estimating volatility cases. Table \ref{comparisonin} shows that the SSE values are 96.3520 of the BS model and 5.1238 for the SV model with implied volatility case and 4.1293 for estimating volatility case. Define averages of error (AE) between the Heston model price $C_{Heston}$ and the market price $C_{market}$ as $|C_{Heston}-C_{market}|$ and averages of relative error (ARE) as $|\frac{C_{Heston}-C_{market}}{C_{market}} |$. The AE and ARE values are remarkably dropping from the BS model to the Heston model for all cases.  Table \ref{comparisonin} also shows that for in-sample data, the SSE, AE and ARE values decreases slightly as 1\%, 2\% and 6\% respectively when jump is considered for the first panel. All the SSE, AE and ARE are improved from the BS model to the Heston model, and from the SV model to the SVJ model for both cases in each corresponding panel, except in panel B the AE and ARE of the SV model for the implied volatility case are smaller than those of the SVJ model, and the ARE of the SV model for the estimating volatility case is smaller than that of the SVJ model. 
\vspace{.15 in}

\textbf{2. Comparison of the BS model and the Non-iid model}

With the models of three cases of non-iid price jumps distribution, we may compare with classic BS model together.

For the case with jumps, the parameters are using market implied volatility or $V_t=0.009$, and $\lambda=0.053$, $\alpha=\mu_J=-0.0001$, $\gamma=\sigma_J=0.0009$ consistently of Panel A in the Heston model calibrations.
The half-year in-sample non-iid cases empirical comparisons against the BS model are showing in Table \ref{noniidin}. 

\begin{table}[]
\small
\raggedright
\caption{Comparison with the BS model and Non-iid Cases for In-sample Data.}
For the first case in the model next to the BS model, take $n=50$, $\mu_1=\mu_2=\mu_3=\mu_J=-0.0001$, $\mu_4=-0.03$, $\mu_5=-0.01$, $\mu_6$ and the others $\mu_i$ are zeroes for the varying mean in the jumping size; the jumping variance $\sigma_i=\sigma_J$ keeps unchanged. 
For the second case of non-iid models with varying variance in the jumping size, take $n=50$, the mean in the jumping size $\mu_i=\mu_J=-0.0001$ is fixed,
and the variance of the jumping size varies  $\sigma_1=\sigma_2=\sigma_3=\sigma_J=0.0009$, $\sigma_4=0.05$, $\sigma_5=0.03$, $\sigma_6=0.01$ and the others are zeroes. 
For the third case of autocorrelated non-iid models, as using the parameters from the Heston model, $n=50$, $\lambda=0.053$, $\alpha=\mu_J=-0.0001$, $\gamma=\sigma_J=0.0009$, $\rho_{il,i\neq l}=0.6891 $. AE means the average of  errors, ARE means the average of relative errors. The percentages in the parenthesis are improvements according to BS model prices.\\
\label{noniidin}
\resizebox{\textwidth}{!}{%
\begin{tabular}{llllllllllllll}
\\
\hline
 &  & BS &  & Cor 1 &  &  & Cor 2 &  &  & Cor 3 (n=50) &  & (n=15) &  \\ \cline{5-6} \cline{8-9} \cline{11-14} 
 &  &  &  & $imp.V_t^2$ & $V_t$  = 0.009 &  & $imp.V_t^2$ & $V_t$  = 0.009 &  & $imp.V_t^2$ & $V_t$  = 0.009 & $imp.V_t^2$ & $V_t$  = 0.009 \\ \cline{3-3} \cline{5-6} \cline{8-9} \cline{11-14} 
SSE &  & 96.3520 &  & 96.3076 & 93.1146 &  & 96.3075 & 93.1146 &  & 96.3076 & 93.1145 & 96.3076 & 93.1146 \\
 &  &  &  & (0.044\%) & (3.358\%) &  & (0.044\%) & (3.358\%) &  & (0.044\%) & (3.358\%) & (0.044\%) & (3.358\%) \\
AE &  & 15.8135 &  & 15.8093 & 15.4337 &  & 15.8093 & 15.4337 &  & 15.8093 & 15.4337 & 15.8093 & 15.4337 \\
 &  &  &  & (0.004\%) & (2.380\%) &  & (0.004\%) & (2.380\%) &  & (0.004\%) & (2.380\%) & (0.004\%) & (2.380\%) \\
ARE &  & 4.4620 &  & 4.4606 & 4.3081 &  & 4.4606 & 4.3081 &  & 4.4606 & 4.3081 & 4.4606 & 4.3081 \\
 &  &  &  & (0.031\%) & (3.449\%) &  & (0.031\%) & (3.449\%) &  & (0.031\%) & (3.449\%) & (0.031\%) & (3.449\%)\\ \hline
\end{tabular}%
}
\end{table}

Table \ref{noniidin} illustrates four models that the BS model against the three non-iid models with mean-varying, variance-varying and auto-correlated non-iid cases. 
For the auto-correlated non-iid case, we set two estimations for $n=50$ and $n=15$. 
For the first case in the model next to the BS model, take $n=50$, $\mu_1=\mu_2=\mu_3=\mu_J=-0.0001$, $\mu_4=-0.03$, $\mu_5=-0.01$, $\mu_6$ and the others $\mu_i$ are zeroes for the varying mean in the jumping size; the jumping variance $\sigma_i=\sigma_J$ keeps unchanged.  We get the results as $SSE=96.3076$ and $SSE=93.1146$ for implied volatility and estimated volatility cases respectively for in-sample. 

For the second case of non-iid models with varying variance in the jumping size, take $n=50$, the mean in the jumping size $\mu_i=\mu_J=-0.0001$ is fixed,
and the variance of the jumping size varies  $\sigma_1=\sigma_2=\sigma_3=\sigma_J=0.0009$, $\sigma_4=0.05$, $\sigma_5=0.03$, $\sigma_6=0.01$ and the others are zeroes. The results are $SSE=96.3075$ for implied volatility and $SSE=93.1146$ for estimated volatility case for in-sample. 

For the third case of autocorrelated non-iid models, as using the parameters from the Heston model, $n=50$, $\lambda=0.053$, $\alpha=\mu_J=-0.0001$, $\gamma=\sigma_J=0.0009$, $\rho_{il,i\neq l}=0.6891 $. The calculation results are $SSE =96.3076$ for implied volatility and $SSE =93.1145$ for estimated volatility for in-sample. 

By checking results in Table \ref{noniidin}, we can see that besides the SSE values, the AE and ARE values of three non-iid cases are very close to the BS model results as $SSE=96.3520$, $15.8135$ and $4.4620$ respectively. 
With the autocorrelation 0.95 between two consecutive jumps and decreases for 5 percent for every next period, we choose $n=15$, then the farthest autocorrelation is $0.3$. The almost same outcomes are obtained as the previous case when $n=50$. 
The improvement percentages are around 3\% for the three non-iid cases in Table \ref{noniidin} when $V_t$ is fixed as $0.009$, and less than 0.1\% for using market implied volatility.

Meanwhile, we want to know the sensitivities for parameters $\alpha$ and $\gamma$. Furthermore, the influence of combined movements in both is tested. We check one day data for simplicity.

For one day data, we analysis 3 cases of the non-iid model: First is to set $\alpha$ increase from $-0.03$, $-0.01$ to $-0.0001$ then stay, and the parameter $\gamma$ is always $0.0009$ unchanged; Second is to set $\alpha$ fixed as $-0.0001$, $\gamma$ decrease from $0.05$, $0.03$ to $0.01$, then stay at $0.0009$; Third case is to let $\alpha$ increase from $-0.03$, $-0.01$ to $-0.0001$ then stays, and simultaneously $\gamma$ decreases from $0.05$, $0.03$ to $0.01$, then stays in $0.0009$. Choosing the first day of the in-samples data then the results are listed in Table \ref{noniid1dayin}.

\begin{table}[]
\small
\raggedright
\caption{Non-iid Cases Analysis of one Day Data for 9/4/2012.(272 prices)} AE means Average Error, ARE means Average Relative Error. The percentages in the parenthesis are improvements according to BS model prices.
For the first day of in-sample data, we analysis 3 cases of the non-iid model: First is to set $\alpha$ increase from $-0.03$, $-0.01$ to $-0.0001$ then stay, and the parameter $\gamma$ is always $0.0009$ unchanged; Second is to set $\alpha$ fixed as $-0.0001$, $\gamma$ decrease from $0.05$, $0.03$ to $0.01$, then stay in $0.0009$; Third case is to let $\alpha$ increase from $-0.03$, $-0.01$ to $-0.0001$ then stays, and simultaneously $\gamma$ decreases from $0.05$, $0.03$ to $0.01$, then stays in $0.0009$. \\
\label{noniid1dayin}
\resizebox{\textwidth}{!}{%
\begin{tabular}{lllllllllllll}
\\
\hline
 &  & BS &  & Cor 1 & Case 1 & Case 2 & Case 3 &  & Cor 2 & Case 1 & Case 2 & Case 3 \\ \cline{3-3} \cline{5-8} \cline{10-13} 
SSE &  & 1.0997 &  & 1.0519 & 1.0614 & 1.0423 & 1.049 &  & 1.0519 & 1.0614 & 1.0426 & 1.0492 \\
 &  &  &  & (4.35\%) & (3.48\%) & (5.22\%) & (4.61\%) &  & (4.35\%) & (3.48\%) & (5.19\%) & (4.59\%) \\
AE &  & 15.7474 &  & 15.2809 & 15.3664 & 15.1907 & 15.2544 &  & 15.2809 & 15.3664 & 15.1929 & 15.2561 \\
 &  &  &  & (2.96\%) & (2.42\%) & (3.54\%) & (3.13\%) &  & (2.96\%) & (2.42\%) & (3.52\%) & (3.12\%) \\
ARE &  & 2.8612 &  & 2.7669 & 2.7961 & 2.7537 & 2.7722 &  & 2.7669 & 2.7961 & 2.7541 & 2.7724 \\
 &  &  &  & (3.30\%) & (2.28\%) & (3.76\%) & (3.11\%) &  & (3.30\%) & (2.28\%) & (3.74\%) & (3.10\%)\\ \hline
\end{tabular}%
}
\end{table}

The improvement percentages show that both case 2 (e.g. for Corollary 1, SSE improvement is 5.22\%) and case 3 (e.g., 4.61\%) are better than the experiments show in Table \ref{noniid1dayin} (e.g., 4.35\%) but the case 1 (e.g., 3.48\%) is worse. And the case 2 has the best improvement over all.  Corollary 1 and Corollary 2 in the Non-iid model have almost the same improvement effects considering $0.03\%$ fluctuation due to the same scenario of these three cases.

\vspace{.15 in}    
\textbf{3. Empirical Test for S\&P 500 Call Option With L\'evy Processes}
     
As the last part of empirical test, three typical L\'evy Processes we mentioned before are checked by optimal parameters searching. The comparison with previous Heston model is also considered.

For the data of S\&P 500 option prices, we get the SSE values for these three L\'evy processes for both in-sample and out-of-sample periods. The first step is to estimate the parameters for in-sample data, which is from September 4th 2012 to February 28th 2013. \cite{gong2016option} use intelligent optimization Differential Evolution algorithm to calibrate the parameters for FFT estimation of L\'evy processes dynamics. In our paper, the task is to make the SSE value be the minimum by using Matlab optimal function "$fmincon$" due to the convenience of software toolbox. 

As the initial parameter values for the optimal searching, results of \cite{schoutens2003levy} which calibrated the S\&P 500 option index are
referred. For the GH model, the initial values are $\alpha=3.8$, $\beta=-3$, $\delta=1$, $\nu=2$; and the NIG model utilize $\alpha=6$, $\beta=-3$, $\mu=0.01$, $\delta=1$; and the CGMY model use $C=0.02$, $G=0.08$, $M=7.55$, $Y=1.3$ for the initial parameters. All the estimations have upper bound $20$ and lower bound $-20$ for the searching regions.

Table \ref{3Levyin} below includes the parameters estimation and the corresponding SSE values.

\begin{table}[]
\small
\raggedright
\caption{SSE Estimations According to 3 Types of L\'evy Processes Comparing Previous Heston Models for In-sample.} The in-sample data comes from September 4th 2012 to February 28th 2013. These three typical L\'evy processes have the equivalent order of magnitudes with the Heston model for the final SSE. 
For the GH model, the initial values are $\alpha=3.8$, $\beta=-3$, $\delta=1$, $\nu=2$; for the CGMY model use $C=0.02$, $G=0.08$, $M=7.55$, $Y=1.3$ for the initial parameters; and for the NIG model utilize $\alpha=6$, $\beta=-3$, $\mu=0.01$, $\delta=1$ for the initial parameters based on results of \cite{schoutens2003levy}. All the estimations have upper bound $20$ and lower bound $-20$ for the searching regions. The greatest in-sample SSE result is 7.57 for GH model and the lowest is 4.13 for NIG model, the average is 6.21. According to Heston model, the greatest is 5.12 and lowest is 4.10 for in-sample with average 4.61. By comparison, the NIG model has the lowest answer in the 3 kinds of L\'evy processes and very close to Heston model.\\
\label{3Levyin}
\resizebox{\textwidth}{!}{%
\begin{tabular}{llllllllllllll}
\\
\hline
 &  & GH &  & NIG &  & CGMY &  & \multicolumn{2}{l}{Heston SV model} &  & \multicolumn{2}{l}{SVJ model} \\ \cline{9-10} \cline{12-13} 
 &  & model &  & model &  & model &  & $imp.V_t^2$ & estimate $V_t$ &  & $imp.V_t^2$ & estimate $V_t$ \\ \cline{3-3} \cline{5-5} \cline{7-7} \cline{9-10} \cline{12-13} 
$\alpha$ &  & -17.3388 &  & 19.999 &  &  &  &  &  &  &  &  \\
$\beta$ &  & 15.6454 &  & -15.421 &  &  &  &  &  &  &  &  \\
$\delta$ &  & 0.3554 &  & 0.2295 &  &  &  &  &  &  &  &  \\
$\nu$ &  & -17.6347 &  &  &  &  &  &  &  &  &  &  \\
$\mu$ &  &  &  & -0.1667 &  &  &  &  &  &  &  &  \\
C &  &  &  &  &  & 4.47E-06 &  &  &  &  &  &  \\
G &  &  &  &  &  & 4.5725 &  &  &  &  &  &  \\
M &  &  &  &  &  & 6.8383 &  &  &  &  &  &  \\
Y &  &  &  &  &  & 1.9975 &  &  &  &  &  &  \\
SSE &  & 6.92 &  & 4.13 &  & 7.57 &  & 5.12 & 4.13 &  & 5.10 & 4.10\\ \hline
\end{tabular}%
}
\end{table}

Table \ref{3Levyin} shows that the biggest in-sample SSE result is 7.57 for GH model and the smallest is 4.13 for NIG model, the average is 6.21.
From the final SSE results, we have that the L\'evy processes have the equivalent order of magnitude of SSE values with the Heston model but slightly larger.  According to the Heston model Panel A (we only consider Panel A for simplicity hereafter), the biggest is 5.12 and the smallest is 4.10 for in-sample with average of 4.61. By comparison, the NIG model has the smallest answers in the three kinds of L\'evy processes and very close to the Heston model.

 \subsection{Out-of-sample Pricing Performance}
 
\vspace{.15 in}    
\textbf{1. Comparison of the BS model and the Heston Model }

For the out-of-sample data, there are total $34,468$ option prices of the S\&P 500 index. By using the parameters of in-sample estimation to evaluate, comparison for BS model and Heston model is in Table \ref{comparisonout}.

\begin{table}[]
\small
\raggedright
\caption{Comparison with the BS and SV, SVJ Models for Out-of-sample Data.}
The out-of-sample period is from March 1, 2013 to August 30, 2013 of the S\&P 500 index option prices. All models are using the parameters from in-sample estimations. The Heston model  decreases the SSE value dramatically for the BS model for both implied volatility case and estimating volatility parameter case.  Besides the SSE values, the Average Error (AE) and ARE means Average Relative Error (ARE) both have noticeable drops from the BS model to the Heston model for all cases. \\
\label{comparisonout}
\resizebox{\textwidth}{!}{%
\begin{tabular}{llllllllllllll}
\\
\hline
 &  &  & \multicolumn{5}{l}{Panel A (upbound=100)} &  & \multicolumn{5}{l}{Panel B (upbound=500)} \\ \cline{4-8} \cline{10-14} 
 & BS &  & SV &  &  & SVJ &  &  & SV &  &  & SVJ &  \\ \cline{4-5} \cline{7-8} \cline{10-11} \cline{13-14} 
 &  &  & $imp.V_t^2$ & estimate $V_t$ &  & $imp.V_t^2$ & estimate $V_t$ &  & $imp.V_t^2$ & estimate $V_t$ &  & $imp.V_t^2$ & estimate $V_t$ \\ \cline{2-2} \cline{4-5} \cline{7-8} \cline{10-11} \cline{13-14} 
SSE & 104.6249 &  & 6.6083 & 5.3216 &  & 6.5496 & 5.2743 &  & 5.5822 & 4.7935 &  & 5.6248 & 4.7672 \\
AE & 17.2347 &  & 2.9657 & 2.6315 &  & 2.9459 & 2.5960 &  & 2.7488 & 2.4067 &  & 2.7577 & 2.3761 \\
ARE & 2.4976 &  & 0.5995 & 0.4717 &  & 0.5856 & 0.4561 &  & 0.4342 & 0.3739 &  & 0.4557 & 0.3747\\ \hline
\end{tabular}%
}
\end{table}

From the comparison, we see the Heston model decreases the SSE value dramatically for the BS model for both implied volatility case and estimating volatility case. The SSE values are 104.6249 of the BS model and 6.6083 for the Heston model of the first case with implied volatility estimates. Besides the SSE values, the Average Error (AE) and Average Relative Error (ARE) both have noticeable drops from the BS model to the Heston model for all cases.
From the SV model to the SVJ model, improvements are 1\%, 1\%, and 2.3\% on SSE, AE and ARE respectively for Panel A. That means the improvement with jumps contributing in calibration with the Heston models is small.

\vspace{.15 in}
\textbf{2. Comparison of the BS model and Non-iid Cases }

With the same parameters setting for in-sample empirical tests, we obtain the out-of-sample results of non-iid jump cases in Table \ref{noniidout}.
The half years of out-of-sample non-iid cases empirical comparisons against the BS model are presented in Table \ref{noniidout}.

\begin{table}[]
\small
\raggedright
\caption{Non-iid Cases Analysis for Out-of-sample Data.}
The non-iid model uses those parameters estimated from in-sample data, and for each Corollary case the parameters are chosen from the Table 5. The out-of-sample empirical test is to have the option market prices of S\&P 500 index from March 1, 2013 to August 30, 2013. For each case of non-iid model, there are two ways to choose the volatility parameter what are the implied volatility and the estimated volatility. Corollary 3 with auto-correlated non-iid is tested with both $n=15$ and $n=50$ for accuracy.
AE means Average Error, ARE means Average Relative Error. The percentages in the parenthesis are improvements according to the BS model prices.\\
\label{noniidout}
\resizebox{\textwidth}{!}{%
\begin{tabular}{llllllllllllll}
\\
\hline
 &  & BS &  & Cor 1 &  &  & Cor 2 &  &  & Cor 3 (n=50) &  & (n=15) &  \\ \cline{5-6} \cline{8-9} \cline{11-14} 
 &  &  &  & $imp.V_t^2$ & $V_t$  = 0.009 &  & $imp.V_t^2$ & $V_t$  = 0.009 &  & $imp.V_t^2$ & $V_t$  = 0.009 & $imp.V_t^2$ & $V_t$  = 0.009 \\ \cline{3-3} \cline{5-6} \cline{8-9} \cline{11-14} 
SSE &  & 104.6249 &  & 104.5680 & 101.0453 &  & 104.5680 & 101.0452 &  & 104.5680 & 101.0453 & 104.5680 & 101.0453 \\
 &  &  &  & (0.050\%) & (3.417\%) &  & (0.050\%) & (3.417\%) &  & (0.050\%) & (3.417\%) & (0.050\%) & (3.417\%) \\
AE &  & 17.2347 &  & 17.2294 & 16.8378 &  & 17.2294 & 16.8378 &  & 17.2294 & 16.8378 & 17.2294 & 16.8378 \\
 &  &  &  & (0.003\%) & (2.276\%) &  & (0.003\%) & (2.276\%) &  & (0.003\%) & (2.276\%) & (0.003\%) & (2.276\%) \\
ARE &  & 2.4976 &  & 2.4967 & 2.4303 &  & 2.4967 & 2.4303 &  & 2.4967 & 2.4303 & 2.4967 & 2.4303 \\
 &  &  &  & (0.036\%) & (2.695\%) &  & (0.036\%) & (2.695\%) &  & (0.036\%) & (2.695\%) & (0.036\%) & (2.695\%)\\ \hline
\end{tabular}%
}
\end{table}

For the first case, we get the SSE results as $104.5680$ for implied volatility and $101.0453$ for estimated volatility cases.  
For the second case, the SSE results are $104.5680$ and $101.0452$ for implied volatility and estimated volatility cases respectively.
For the third case, the estimation SSE results are $101.5680$ for implied volatility and $101.0453$ for estimated volatility cases when $n=50$. 
Including the checking of AE and ARE values for out-of-sample, all the values are very close to the BS model results with only slight improvement.

Optional case for choosing $n=15$ according to the in-sample test, the same outcomes are obtained for out-of-sample with the previous case as $n=50$.
Both comparisons of in-sample and out-of-sample of non-iid cases suggest the fact that the non-iid assumption for the BS model only improve the performance of misspecific calibration very slightly.

Consistently, sensitivities for parameters $\alpha$ and $\gamma$ and the influence of combined movements in both are tested. 
Choosing the first days of out-of-sample for simplicity then the results are listed in Table \ref{noniid1dayout}.
The improvement percentages keep the same change trends as we list in previous in-sample test. The case in Table \ref{noniid1dayout} is the same case in
Table 6 for the first day of in-sample data on varying the jumping size mean, variance and both.

\begin{table}[]
\small
\raggedright
\caption{Non-iid Cases Analysis of one Day Data for 3/1/2013.(187 prices)}
For the first day of out-of-sample data, we analysis 3 cases of the non-iid model: First is to set $\alpha$ increase from $-0.03$, $-0.01$ to $-0.0001$ then stay, and the parameter $\gamma$ is always $0.0009$ unchanged; Second is to set $\alpha$ fixed as $-0.0001$, $\gamma$ decrease from $0.05$, $0.03$ to $0.01$, then stay in $0.0009$; Third case is to let $\alpha$ increase from $-0.03$, $-0.01$ to $-0.0001$ then stays, and simultaneously $\gamma$ decreases from $0.05$, $0.03$ to $0.01$, then stays in $0.0009$. The corresponding cases are exactly as same as those for the first day of in-sample data in Table 6. AE means Average Error, ARE means Average Relative Error. The percentages in the parenthesis are improvements according to BS model prices.\\
\label{noniid1dayout}
\resizebox{\textwidth}{!}{%
\begin{tabular}{lllllllllllll}
\\
\hline
 &  & BS &  & Cor 1 & Case 1 & Case 2 & Case 3 &  & Cor 2 & Case 1 & Case 2 & Case 3 \\ \cline{3-3} \cline{5-8} \cline{10-13} 
SSE &  & 0.5747 &  & 0.5572 & 0.5623 & 0.5521 & 0.5551 &  & 0.5572 & 0.5623 & 0.5523 & 0.5553 \\
 &  &  &  & (3.05\%) & (2.16\%) & (3.93\%) & (3.41\%) &  & (3.05\%) & (2.16\%) & (3.90\%) & (3.38\%) \\
AE &  & 16.2018 &  & 15.8389 & 15.9257 & 15.7393 & 15.7959 &  & 15.8389 & 15.9257 & 15.7425 & 15.7984 \\
 &  &  &  & (2.24\%) & (1.70\%) & (2.85\%) & (2.51\%) &  & (2.24\%) & (1.70\%) & (2.83\%) & (2.49\%) \\
ARE &  & 4.5028 &  & 4.3612 & 4.4041 & 4.345 & 4.3655 &  & 4.3612 & 4.4041 & 4.3461 & 4.3663 \\
 &  &  &  & (3.14\%) & (2.19\%) & (3.50\%) & (3.05\%) &  & (3.14\%) & (2.19\%) & (3.48\%) & (3.03\%)\\ \hline
\end{tabular}%
}
\end{table}

In summary, the non-iid  model have no dramatic improvement of the BS model since the non-iid formulas are derived from the classic BS model with compound Poisson without uniform distributions. The underlying dynamic of of the asset is essentially same with only altering the i.i.d in the jumping size.
That is why we should not expect too much improvement from the previous SVJ model.

\vspace{.15in}
\textbf{3. Empirical Test for S\&P 500 Call Option With L\'evy Processes}

From the in-sample estimations of parameters of three typical L\'evy models, we empirically test the European call option price of S\&P 500 index
for the out-of-sample period.
Table \ref{3Levyout} includes the parameter estimation and the corresponding SSE values for the option prices.

\begin{table}[]
\small
\raggedright
\caption{SSE Estimations According to 3 Types of L\'evy Processes Comparing Previous Heston Models for Out-of-sample.} The out-of-sample data comes from March 1st 2013 to August 30th 2013. These three L\'evy processes have the equivalent order of magnitudes with Heston model for the final SSE. The greatest out-of-sample SSE estimation is 8.15 for CGMY model and the lowest is 5.81 for NIG model, the average is 7.13. According to Heston model, the greatest is 6.61 and lowest is 5.27 for out-of-sample with average 5.94. By comparison, and the NIG model has the lowest answer in the 3 kinds of L\'evy processes and very close to Heston model as well.\\
\label{3Levyout}
\resizebox{\textwidth}{!}{%
\begin{tabular}{llllllllllllll}
\\
\hline
&  & GH &  & NIG &  & CGMY &  & \multicolumn{2}{l}{Heston SV model} &  & \multicolumn{2}{l}{SVJ model} \\ \cline{9-10} \cline{12-13} 
 &  & model &  & model &  & model &  & $imp.V_t^2$ & estimate $V_t$ &  & $imp.V_t^2$ & estimate $V_t$ \\ \cline{3-3} \cline{5-5} \cline{7-7} \cline{9-10} \cline{12-13} 
$\alpha$ &  & -17.3388 &  & 19.999 &  &  &  &  &  &  &  &  \\
$\beta$ &  & 15.6454 &  & -15.421 &  &  &  &  &  &  &  &  \\
$\delta$ &  & 0.3554 &  & 0.2295 &  &  &  &  &  &  &  &  \\
$\nu$ &  & -17.6347 &  &  &  &  &  &  &  &  &  &  \\
$\mu$ &  &  &  & -0.1667 &  &  &  &  &  &  &  &  \\
C &  &  &  &  &  & 4.47E-06 &  &  &  &  &  &  \\
G &  &  &  &  &  & 4.5725 &  &  &  &  &  &  \\
M &  &  &  &  &  & 6.8383 &  &  &  &  &  &  \\
Y &  &  &  &  &  & 1.9975 &  &  &  &  &  &  \\
SSE  &  & 7.43 &  & 5.81 &  & 8.15 &  & 6.61 & 5.32 & & 6.55 & 5.27\\ \hline
\end{tabular}%
}
\end{table}

Table \ref{3Levyout} illustrates that the biggest out-of-sample SSE estimation is 8.15 of CGMY model and the smallest is 5.81 of NIG model, the average is 7.13. For the Heston models, the biggest is 6.61 and the smallest is 5.27 for out-of-sample with average of 5.94. All those parameters for the GH, NIG, CGMY models are derived from minimizing the SSE functions with proper ranges of each parameter. The SV and SVJ models with estimating volatility give the smallest SSE values 5.32 and 5.27 respectively, where the NIG model follows next with 5.81 SSE value. 

By comparing these L\'evy models with SV and SVJ models, L\'evy processes have the equivalent order of magnitude of SSE values with Heston model for out-of-sample test, and the NIG model performances the lowest answers among the three typical  L\'evy processes and very close to the Heston model. This fact verifies the viewpoint of \cite{geman2001time} about the infinite arriving jump may represent the diffusion part as Heston model deals with. Furthermore, we have some stable features in the prediction of L\'evy processes in the following subsection.

\subsection{Two Scenario Calibrations for One Day Prediction of L\'evy Processes}
To detect the prediction of L\'evy processes, another test about daily SSE computation is undertaken. We use 273 option prices of September 4th 2012 as in-sample data and 187 prices of the first day of second half year, March 1st 2013, as out-of-sample data. With the first day parameters, the out-of-sample results are listed in Table \ref{3Levy1dayout} as well. The SSE values are computed under two scenarios: using the first day parameters and using the first half year parameters. 

Table \ref{3Levy1dayout} shows the parameter results of three L\'evy processes for in-sample case. 
 \begin{table}[]
\small
\caption{Estimation about One Day of Out-of-sample, March 1st 2013, in Three L\'evy Processes.} 
\flushleft
For each model, the first column is the parameters for September 4th 2012 optimization and the second column lists the parameters of half year in-sample calibration.
For the first column of GH model, the SSE value in September 4th 2012 is 0.0415, and the SSE value for March 1st 2013 is 0.0845. The results of out-of-sample by using one day calibration parameters are more than double of in-sample SSE values. On the other side, the SSE value for GH model on second column is 0.0277, which means by using the first half year's optimal parameters, the value becomes almost one quarter. And the CGMY model does as the same. The effect of NIG model is as dramatic as around fifteen-fold less since the SSE value changes from 0.2019 to 0.0139.

\label{3Levy1dayout}
\resizebox{\textwidth}{!}{%
\begin{tabular}{lllllllllll}
\\
\hline
 &  &  & GH model &  &  & NIG model &  &  & CGMY model &  \\ \cline{4-5} \cline{7-8} \cline{10-11} 
$\alpha$ &  &  & 3.1068 & -17.3388 &  & 9.2051 & 19.999 &  &  &  \\
$\beta$ &  &  & 2.0105 & 15.6454 &  & -9.2051 & -15.421 &  &  &  \\
$\delta$ &  &  & 0.4748 & 0.3554 &  & 0.3160 & 0.2295 &  &  &  \\
$\nu$ &  &  & -18.3818 & -17.6347 &  &  &  &  &  &  \\
$\mu$ &  &  &  &  &  & -1.1726 & -0.1667 &  &  &  \\
C &  &  &  &  &  &  &  &  & 4.72E-06 & 4.47E-06 \\
G &  &  &  &  &  &  &  &  & 1.4314 & 4.5725 \\
M &  &  &  &  &  &  &  &  & 7.2307 & 6.8383 \\
Y &  &  &  &  &  &  &  &  & 1.9984 & 1.9975 \\
SSE & Sept. 4th 2012 &  & 0.0415 &  &  & 0.0530 &  &  & 0.0453 &  \\
 & Mar. 1st 2013 &  & 0.0845 & 0.0277 &  & 0.2019 & 0.0139 &  & 0.0852 & 0.0288\\ \hline
\end{tabular}%
}
\end{table}

For each model, table \ref{3Levy1dayout} consists of the first column that is the parameters for September 4th 2012 optimization, and the second column that lists the parameters of half year in-sample calibration.
For the first column of the GH model, the SSE value in September 4th 2012 is
0.0415, and the SSE value for March 1st 2013 is 0.0845. The results of
out-of-sample by using one day calibration parameters are more than double of in-sample SSE values. On the other side, the SSE value for the GH model on
the second column is 0.02774, which means by using the first half year's optimal
parameters, the value becomes almost one quarter of the SSE value for March 1st 2013. The CGMY model does the same on the SSE. The effect of the NIG model is as dramatic as around fifteen-fold less since the SSE value changes from 0.2019 to 0.0139. Overall the NIG model based on the first half year's optimal parameters performs the smallest SSE, and the CGMY based on September 4th 2012 optimization performs the biggest SSE for the one day of out-of-sample test.

The change of SSE values indicates that the prediction error should be narrowed by using previous long period calibration parameters, not short period calibration as the one day optimal parameters.   
One more unexpected thing  is that there are some negative values for the GH model estimation. By checking the model prices of the GH model, 42 negative values happened in the 272 prices of first day data, and the reason for the negative values comes from the iterating procedure. We further investigate all of these negative model prices are corresponding to very small market prices of S\&P 500 index option. 
When we filter and keep all the positive 230 prices for the first day data, we notice the SSE value drops to 0.0352 from 0.0415 in Table \ref{3Levy1dayout}. Comparing results after filtering, both the total prices and SSE values drop 15\% coincidentally.
For the CGMY model and NIG model, we find no negative values for the model price estimation. In all, the unexpected happening about negative estimation does not affect the analyze for the total consideration.
For out-of-sample data, we have 181 positive model prices in GH model from 187 items and the SSE value changes from 0.0845 in table \ref{3Levy1dayout} to 0.0836. The change percents of total prices and SSE are 3\% and 1\% respectively, changes are not different dramatically. 

Intuitively, we may compare daily prediction results with previous Heston model.
For the daily calibration, the parameter of Heston model are estimated by Matlab optimal function and listed in Table \ref{Heston1daypara}.
With Heston models parameters, Table \ref{3Levy1dayout} and \ref{Heston1daypara} show the results and comparison of L\'evy processes with previous Heston model.

\begin{table}[]
\flushleft
\small
\caption{One Day SSE Calibration of Heston Model for Comparison with Three Types of L\'evy processes.} 
For each model and the case of value $V_t$, the first two columns include the parameters for September 4th 2012 optimization and the second two column list the parameters of half year in-sample calibration.
Checking these cases of Heston model, the SSE values show some facts of oscillation. Extraordinarily, the fifth case of SSE value is 7.4027 for using first day estimation parameters, which looks protruding large. However, the three L\'evy processes have the same shrink trend when we use the long period data calibration parameters. For example, the NIG model has the change from 0.2019 to 0.0139 and lower than any Heston model. Which means The L\'evy processes have the more stable prediction which is expectable with less error for chosen data.
\label{Heston1daypara}
\resizebox{\textwidth}{!}{%
\begin{tabular}{llllllllllllll}
\\
\hline
 &  &  & \multicolumn{5}{l}{SV} &  & \multicolumn{5}{l}{SVJ} \\ \cline{4-5} \cline{7-8} \cline{10-11} \cline{13-14} 
$\kappa$ &  &  & 19.9998 & 20.0000 &  & 4.7985 & 12.8826 &  & 4.7747 & 19.9994 &  & 5.9353 & 13.2477 \\
$\theta$ &  &  & 0.0546 & 0.0112 &  & 0.0062 & 0.0125 &  & 0.3220 & 0.0109 &  & 0.0073 & 0.0131 \\
$\sigma$ &  &  & 2.0127 & 0.7329 &  & 0.1664 & 0.1399 &  & 2.4965 & 0.7244 &  & 0.1242 & 0.1551 \\
$\rho$ &  &  & -0.0172 & 0.1954 &  & 0.9984 & 0.7938 &  & 0.4356 & 0.1921 &  & 0.9984 & 0.6891 \\
$V_t$ &  &  & $imp.V_t^2$ & $imp.V_t^2$ &  & 0.0090 & 0.0089 &  & $imp.V_t^2$ & $imp.V_t^2$ &  & 0.0098 & 0.009 \\
$\lambda$ &  &  &  &  &  &  &  &  & 0.0034 & 0.5207 &  & 0.00016 & 0.0530 \\
$\mu_j$ &  &  &  &  &  &  &  &  & 0.5415 & 0.0001 &  & -0.0518 & -0.0001 \\
$\sigma_j$ &  &  &  &  &  &  &  &  & 1.2229 & 0.0002 &  & 1.0009 & 0.0009 \\
SSE & Sept. 4th 2012 &  & 0.0431 &  &  & 0.0059 &  &  & 0.0366 &  &  & 0.0068 &  \\
 & Mar. 1st 2013 &  & 0.0271 & 0.0278 &  & 0.0857 & 0.0202 &  & 7.4027 & 0.0484 &  & 0.1025 & 0.0807\\ \hline
\end{tabular}%
}
\end{table}

Checking the four cases of the Heston model, the SSE values show oscillating behavior without clear pattern. For the SV model in the first case when we use implied volatility, the estimation of out-of-sample by using first day parameters has 0.0271 SSE value, and the SSE value becomes a slightly bigger 0.0278 for using first half year parameters. For the second case when we use estimated volatility, the SSE value varies from 0.0857 to 0.0202 with a drop. Extraordinarily, the SSE value in the third case of SVJ model is 7.4027 for using first day estimation parameters, which looks protrudingly large. However, three L\'evy processes have the same shrinking trend when we use the long period data calibration parameters. For example, the NIG model has the change from 0.02019 to 0.0139 and lower than any Heston model we used. This indicates that L\'evy processes have the more stable prediction with less errors as we expected. 

In summary, the SSE values of L\'evy processes have the same order of magnitude with Heston model we did before and even slightly greater. 
However,  the one day prediction for out-of-sample data has a
shrinking effect for the L\'evy processes and more stable than Heston model. In articular, the NIG model has the lowest SSE value prediction within all the models.

\section{Conclusion}

Brownian motion is omnipresent in the financial economics, especially  as  
a  fundamental tool to simulate the dynamics of asset price movement over time. In the field of quantitative finance, the Black-Scholes-Merton model studies the option pricing formula that the underlying asset price follows geometric Brownian motion. 
 
We follow the method of \cite{bakshi1997empirical} to first estimate parameters by the sum of squared error (SSE) for in-sample period of S\&P 500 index. Other than the compound Poisson jumps, we also test the SSE results of non-iid cases numerically. The Heston model provides an excellent improvement of the BSM model which has fixed volatility. The advanced Heston models (SV, SVJ), which consider the volatility is a stochastic process, have the SSE value less than 10. By comparison, we find that the BSM model has the least performance consistently, and SV and SVJ models with estimated volatility have better results than SV and SVJ models with implied volatility by comparing those models with different  periods of \cite{bakshi1997empirical}.  
 
  Generally, the price activity has jumps which can be observed in the real financial market. L\'evy processes are outstanding in terms of financial mathematics since their infinitely divisible, independent and stationary increments properties match financial market intuitively. For the model of exponential of L\'evy process, the PIDE can be derived for the option pricing formula, but it is very hard to solve in a closed-form theoretically. Nevertheless, Fourier transform method can be used to solve the PIDE numerically due to the analogy of L\'evy process' characteristic function by L\'evy-Khinchin theorem. In this paper, three typical cases of L\'evy processes, GH model, NIG model and CGMY model, are calibrated by using fast Fourier transform (FFT) method numerically. Not only the parameter sensitivities of these L\'evy processes are analyzed, but also  all the SSE results of half year data are evaluated (below 10 as Heston model does). As a result, the NIG model has the better result than the GH and CGMY models for both in-sample and out-of-sample periods consistently. Meanwhile, for in-sample data, both NIG, SV and SVJ with estimated volatility models outperform the rest in terms of the least SSE value. For out-of-sample period, the SV and SVJ models with estimated volatility outperform the rest models and the NIG model follows next. Additionally, for one day prediction on March 1st 2013, we find that the NIG model is the best among all the models with the in-sample optimized parameters. Furthermore, the L\'evy processes have the shrink effect and more stable prediction for chosen data. Beyond those analysis, the hedging problem with respect to the jumps is left for further study to investigate.


\bibliographystyle{jfe}
\bibliography{SP500Levy}

\begin{thebibliography}{49}
\expandafter\ifx\csname natexlab\endcsname\relax\def\natexlab#1{#1}\fi

\bibitem[{Amin and Ng(1993)}]{amin1993option}
Amin, K.~I., Ng, V.~K., 1993. Option valuation with systematic stochastic
  volatility. The Journal of Finance 48, 881--910.

\bibitem[{Bakshi et~al.(1997)Bakshi, Cao, and Chen}]{bakshi1997empirical}
Bakshi, G., Cao, C., Chen, Z., 1997. Empirical performance of alternative
  option pricing models. The Journal of Finance 52, 2003--2049.

\bibitem[{Bakshi and Chen(1997)}]{bakshi1997alternative}
Bakshi, G.~S., Chen, Z., 1997. An alternative valuation model for contingent
  claims. Journal of Financial Economics 44, 123--165.

\bibitem[{Barndorff-Nielsen(1977)}]{barndorff1977exponentially}
Barndorff-Nielsen, O., 1977. Exponentially decreasing distributions for the
  logarithm of particle size. Proceedings of the Royal Society of London. A.
  Mathematical and Physical Sciences 353, 401--419.

\bibitem[{Barndorff-Nielsen and Blaesild(1981)}]{barndorff1981hyperbolic}
Barndorff-Nielsen, O., Blaesild, P., 1981. Hyperbolic distributions and
  ramifications: Contributions to theory and application. In: {\em Statistical
  distributions in scientific work\/}, Springer, pp. 19--44.

\bibitem[{Barndorff-Nielsen(1997)}]{barndorff1997processes}
Barndorff-Nielsen, O.~E., 1997. Processes of normal inverse gaussian type.
  Finance and Stochastics 2, 41--68.

\bibitem[{Bates(1991)}]{bates1991crash}
Bates, D.~S., 1991. The crash of {'}87: Was it expected? the evidence from
  options markets. The Journal of Finance 46, 1009--1044.

\bibitem[{Bates(1996)}]{bates1996jumps}
Bates, D.~S., 1996. Jumps and stochastic volatility: Exchange rate processes
  implicit in deutsche mark options. The Review of Financial Studies 9,
  69--107.

\bibitem[{Black and Scholes(1973)}]{black1973pricing}
Black, F., Scholes, M., 1973. The pricing of options and corporate liabilities.
  Journal of Political Economy 81, 637--654.

\bibitem[{Blaesild and S{\o}rensen(1992)}]{blaesild1992hyp}
Blaesild, P., S{\o}rensen, M., 1992. " Hyp": A Computer Program for Analyzing
  Data by Means of the Hyperbolic Distribution. University of Aarhus,
  Department of Theoretical Statistics.

\bibitem[{Broadie et~al.(2009)Broadie, Chernov, and
  Johannes}]{broadie2009understanding}
Broadie, M., Chernov, M., Johannes, M., 2009. Understanding index option
  returns. The Review of Financial Studies 22, 4493--4529.

\bibitem[{C\^amara and Li(2008)}]{camara2008jump}
C\^amara, A., Li, W., 2008. Jump-diffusion option pricing without iid jumps.
  Available at SSRN: https://ssrn.com/abstract=1282882 or
  http://dx.doi.org/10.2139/ssrn.1282882 .

\bibitem[{Carr et~al.(2002)Carr, Geman, Madan, and Yor}]{carr2002fine}
Carr, P., Geman, H., Madan, D.~B., Yor, M., 2002. The fine structure of asset
  returns: An empirical investigation. The Journal of Business 75, 305--332.

\bibitem[{Carr and Wu(2004)}]{carr2004time}
Carr, P., Wu, L., 2004. Time-changed l\'evy processes and option pricing.
  Journal of Financial Economics 71, 113--141.

\bibitem[{Christoffersen et~al.(2009)Christoffersen, Heston, and
  Jacobs}]{christoffersen2009shape}
Christoffersen, P., Heston, S., Jacobs, K., 2009. The shape and term structure
  of the index option smirk: Why multifactor stochastic volatility models work
  so well. Management Science 55, 1914--1932.

\bibitem[{Constantinides et~al.(2008)Constantinides, Jackwerth, and
  Perrakis}]{constantinides2008mispricing}
Constantinides, G.~M., Jackwerth, J.~C., Perrakis, S., 2008. Mispricing of s\&p
  500 index options. The Review of Financial Studies 22, 1247--1277.

\bibitem[{Cont and Tankov(2003)}]{Cont2003financial}
Cont, R., Tankov, P., 2003. Financial modelling with jump processes, vol.~2.
  CRC press.

\bibitem[{Cont and Voltchkova(2005)}]{cont2005integro}
Cont, R., Voltchkova, E., 2005. Integro-differential equations for option
  prices in exponential l\'evy models. Finance and Stochastics 9, 299--325.

\bibitem[{Cooley and Tukey(1965)}]{cooley1965algorithm}
Cooley, J.~W., Tukey, J.~W., 1965. An algorithm for the machine calculation of
  complex fourier series. Mathematics of Computation 19, 297--301.

\bibitem[{Cox et~al.(1985)Cox, Ingersoll~Jr, and Ross}]{cox1985theory}
Cox, J.~C., Ingersoll~Jr, J.~E., Ross, S.~A., 1985. A theory of the term
  structure of interest rates. Econometrica 53, 385--408.

\bibitem[{Duffie et~al.(2000)Duffie, Pan, and Singleton}]{duffie2000transform}
Duffie, D., Pan, J., Singleton, K., 2000. Transform analysis and asset pricing
  for affine jump-diffusions. Econometrica 68, 1343--1376.

\bibitem[{Florescu et~al.(2014)Florescu, Mariani, and
  Sewell}]{florescu2014numerical}
Florescu, I., Mariani, M.~C., Sewell, G., 2014. Numerical solutions to an
  integro-differential parabolic problem arising in the pricing of financial
  options in a levy market. Quantitative Finance 14, 1445--1452.

\bibitem[{Geman et~al.(2001)Geman, Madan, and Yor}]{geman2001time}
Geman, H., Madan, D.~B., Yor, M., 2001. Time changes for l\'evy processes.
  Mathematical Finance 11, 79--96.

\bibitem[{Gong and Zhuang(2016)}]{gong2016option}
Gong, X., Zhuang, X., 2016. Option pricing for stochastic volatility model with
  infinite activity l\'evy jumps. Physica A: Statistical Mechanics and its
  Applications 455, 1--10.

\bibitem[{Hansen(1982)}]{hansen1982large}
Hansen, L.~P., 1982. Large sample properties of generalized method of moments
  estimators. Econometrica: Journal of the Econometric Society pp. 1029--1054.

\bibitem[{He and Zhu(2018)}]{he2018closed}
He, X.-J., Zhu, S.-P., 2018. A closed-form pricing formula for european options
  under the heston model with stochastic interest rate. Journal of
  Computational and Applied Mathematics 335, 323--333.

\bibitem[{Heston(1993)}]{heston1993closed}
Heston, S.~L., 1993. A closed-form solution for options with stochastic
  volatility with applications to bond and currency options. The Review of
  Financial Studies 6, 327--343.

\bibitem[{Hirsa(2016)}]{hirsa2016computational}
Hirsa, A., 2016. Computational methods in finance. CRC Press.

\bibitem[{Hull and White(1987)}]{hull1987pricing}
Hull, J., White, A., 1987. The pricing of options on assets with stochastic
  volatilities. The Journal of Finance 42, 281--300.

\bibitem[{Kienitz and Wetterau(2012)}]{kienitz2012financial}
Kienitz, J., Wetterau, D., 2012. Financial modelling: Theory, implementation
  and practice with MATLAB source. John Wiley \& Sons.

\bibitem[{Kwok et~al.(2012)Kwok, Leung, and Wong}]{kwok2012efficient}
Kwok, Y.~K., Leung, K.~S., Wong, H.~Y., 2012. Efficient options pricing using
  the fast fourier transform. In: {\em Handbook of computational finance\/},
  Springer, pp. 579--604.

\bibitem[{Lee and Hannig(2010)}]{lee2010detecting}
Lee, S.~S., Hannig, J., 2010. Detecting jumps from l\'evy jump diffusion
  processes. Journal of Financial Economics 96, 271--290.

\bibitem[{Lewis(2000)}]{lewisoption}
Lewis, A., 2000. Option valuation under stochastic volatility with mathematica
  code.

\bibitem[{Li et~al.(2006)Li, Wells, and Yu}]{li2006bayesian}
Li, H., Wells, M.~T., Yu, C.~L., 2006. A bayesian analysis of return dynamics
  with l\'evy jumps. The Review of Financial Studies 21, 2345--2378.

\bibitem[{Li and Li(2015)}]{li2015hedging}
Li, O.~X., Li, W., 2015. Hedging jump risk, expected returns and risk premia in
  jump-diffusion economies. Quantitative Finance 15, 873--888.

\bibitem[{Lord et~al.(2008)Lord, Fang, Bervoets, and Oosterlee}]{lord2008fast}
Lord, R., Fang, F., Bervoets, F., Oosterlee, C.~W., 2008. A fast and accurate
  fft-based method for pricing early-exercise options under l\'evy processes.
  SIAM Journal on Scientific Computing 30, 1678--1705.

\bibitem[{Loretan and Phillips(1994)}]{loretan1994testing}
Loretan, M., Phillips, P.~C., 1994. Testing the covariance stationarity of
  heavy-tailed time series: An overview of the theory with applications to
  several financial datasets. Journal of Empirical Finance 1, 211--248.

\bibitem[{Lux(2000)}]{lux2000moment}
Lux, T., 2000. On moment condition failure in german stock returns: an
  application of recent advances in extreme value statistics. Empirical
  Economics 25, 641--652.

\bibitem[{Merton(1976)}]{merton1976option}
Merton, R.~C., 1976. Option pricing when underlying stock returns are
  discontinuous. Journal of Financial Economics 3, 125--144.

\bibitem[{Merton et~al.(1973)}]{merton1973theory}
Merton, R.~C., et~al., 1973. Theory of rational option pricing. Theory of
  Valuation pp. 229--288.

\bibitem[{Ornthanalai(2014)}]{ornthanalai2014levy}
Ornthanalai, C., 2014. Levy jump risk: Evidence from options and returns.
  Journal of Financial Economics 112, 69--90.

\bibitem[{Rouah(2013)}]{rouah2013heston}
Rouah, F.~D., 2013. The Heston Model and Its Extensions in Matlab and C. John
  Wiley \& Sons.

\bibitem[{Schoutens(2003)}]{schoutens2003levy}
Schoutens, W., 2003. L\'evy processes in finance: pricing financial
  derivatives. John Wiley \& Sons.

\bibitem[{Scott(1987)}]{scott1987option}
Scott, L.~O., 1987. Option pricing when the variance changes randomly: Theory,
  estimation, and an application. Journal of Financial and Quantitative
  Analysis 22, 419--438.

\bibitem[{Scott(1997)}]{scott1997pricing}
Scott, L.~O., 1997. Pricing stock options in a jump-diffusion model with
  stochastic volatility and interest rates: Applications of fourier inversion
  methods. Mathematical Finance 7, 413--426.

\bibitem[{Shreve(2004)}]{shreve2004stochastic}
Shreve, S.~E., 2004. Stochastic calculus for finance II: Continuous-time
  models, vol.~11. Springer Science \& Business Media.

\bibitem[{Strang(1994)}]{strang1994wavelets}
Strang, G., 1994. Wavelets. American Scientist 82, 250--255.

\bibitem[{Wong and Guan(2011)}]{wong2011fft}
Wong, H.~Y., Guan, P., 2011. An fft-network for l\'evy option pricing. Journal
  of Banking \& Finance 35, 988--999.

\bibitem[{Zaevski et~al.(2014)Zaevski, Kim, and Fabozzi}]{zaevski2014option}
Zaevski, T.~S., Kim, Y.~S., Fabozzi, F.~J., 2014. Option pricing under
  stochastic volatility and tempered stable l\'evy jumps. International Review
  of Financial Analysis 31, 101--108.

\end{thebibliography}

\end{document}